\documentclass[12pt,technote,draftclsnofoot,onecolumn]{IEEEtran}

\usepackage{dsfont}
\usepackage{amsmath}
\usepackage{amssymb}
\usepackage{amsfonts}
\usepackage{graphicx}
\usepackage{amsmath}
\usepackage[all,poly]{xy}
\usepackage{multirow}
\usepackage{algorithm}
\usepackage{algorithmic}
\usepackage{color}
\usepackage[noadjust]{cite}
\usepackage{tikz,pgf}
\usetikzlibrary{fit}
\usetikzlibrary{arrows,automata}
\usepackage{verbatim}
\usetikzlibrary{positioning,shapes.geometric}
\usepackage{url}
\usepackage[T1]{fontenc}
\usepackage{subfig}
\usepackage{cleveref}

\newcounter{algo}
\renewcommand{\thealgo}{\arabic{algo}}

\title{Statistical modeling and classification of reflectance confocal microscopy images}

\begin{document}




%
%
\author{Abdelghafour Halimi\thanks{This work was funded by Pierre Fabre Dermo Cosm\'{e}tique.}, Hadj Batatia,  Jimmy Le Digabel, Gwendal Josse and Jean-Yves Tourneret}

 \maketitle
\bigskip
\begin{center} \textbf{\textrm{TECHNICAL REPORT -- 2017, April}}\\
University of Toulouse, IRIT/INP-ENSEEIHT \\2 rue Camichel, BP 7122,
31071 Toulouse cedex 7, France
  \end{center}
\bigskip
\bigskip\bigskip

%
%

\begin{abstract}
This paper deals with the characterization and classification of reflectance confocal microscopy images of human skin. The aim is to identify and characterize the lentigo, a phenomenon that originates at the dermo-epidermic junction of the skin. High resolution confocal images are acquired at different skin depths and are analyzed for each depth. Histograms of pixel intensities associated with a given depth are determined, showing that the generalized gamma distribution (GGD) is a good statistical model for confocal images. A GGD is parameterized by translation, scale and shape parameters. These parameters are estimated using a new estimation method based on a natural gradient descent showing fast convergence properties with respect to state-of-the-art estimation methods. The resulting parameter estimates can be used to classify clinical images of healthy and lentigo patients.  The obtained results show that the scale and shape parameters are  good features to identify and characterize the presence of lentigo in skin tissues. 

\end{abstract}

\section{Introduction} \label{sec:Introduction}
The human skin is a large and complex organ that can be subjected to a number of diseases. The lentigo is a type of lesion that originates at the junction between the dermis and the epidermis due to a high concentration of melanocytes aggregating at the dermal papillae walls. This lesion leads to the destruction of the regular cellular network at certain layers of the epidermis, mainly at the dermoepidermal junction \cite{menge2016concordance}. The diagnosis of lentigo can be performed through visual inspection or through biopsy of the skin surface. Reflectance confocal microscopy (RCM) is a non-invasive imaging technique that enables in vivo visualisation of the epidermis down to the papillary dermis in real time \cite{Nehal2008,Hofmann2009}. Its potential to improve the detection of cancer and tumors has been demonstrated in various research and dermatological clinical studies \cite{Alarconcar2014}. Current practices to analyze these images are mainly based on visual inspection. In \cite{menge2016concordance}, a correlation between RCM and histology has been reported for the diagnosis of melanoma. RCM has also proved valuable for treatment follow up \cite{Alarcon2014}, surveillance of lentigo malign treatment \cite{Guitera2014,champin2014vivo},  and guidance of cutaneous surgery \cite{Hibler2015}.\\

The complex nature of RCM images requires automatic image processing methods to build accurate diagnosis strategies. The literature does not report many of such automatic techniques. Subsequent research on RCM images has mainly focused on three aspects: i) clinical studies to evaluate their usefulness, ii) segmentation of nuclei, and iii) classification of skin tissues. Luck et al. \cite{Luck2005} have first developed an automatic RCM image processing method to segment nuclei. Their method was based on a Gaussian image model that takes into account the reflectivity of nuclei and a truncated Gaussian distribution to represent the intensity of the cytoplasm fibers. A Gaussian Markov random field was also used for spatial correlation, and a Bayesian classification algorithm was investigated to label tissues. Harris et al. \cite{harris2015pulse} developed an algorithm based on neural networks to segment nuclei
in RCM images. Nuclei and background, from manually segmented images,
were represented using Gaussian distributions. The neural network was trained
using features such as nuclear size, contrast, and density. Various features extracted from RCM images have been used for the applications cited above.
Kurugol et al. \cite{kurugol2011semi,kurugol2012validation} developed and validated a semi-automatic method to locate the dermoepidermal junction (DEJ) using a statistical classifier based on texture features. These features are mainly related to brightness measurements associated with basal cells.
The authors of \cite{Somoza2014} developed an automatic method to localize skin layers in RCM images based on texture analysis.
Hames et al. \cite{hames2015anatomical,hames2016automated} developed a logistic regression classifier to automatically segment the different layers of the skin in RCM images. In \cite{kose2016machine}, an SVM classification method based on SURF texture features was used to identify skin morphological patterns in RCM images. Finally, for diagnosis applications, Koller et al. \cite{koller2011vivo} investigated a wavelet-based decision tree classification method to distinguish benign and malignant melanocytic skin tumors in RCM images. This method, will be used as a benchmark in our study.

Very few research works reported in the literature have focused on determining quantitative markers for tissue characterization in RCM images. Among these works, Raphael et al. \cite{raphael2013computational} reported a characterization method of RCM images to assess photoageing. In their study, the intensity, 2D wavelet coefficients, 2D Fourier coefficients and  shapes, segmented with an adhoc algorithm, were correlated with clinical data. The results obtained in \cite{raphael2013computational} showed that the image intensity and the wavelet coefficients have no significant correlation, contrary to Fourier coefficients and segmentation results.

The first contribution of this paper is a statistical model that allows the characterization of the underlying tissues. The variability of the pixel intensities of an RCM image is represented by a GGD, whose parameters are used as features for the classification of healthy and lentigo confocal images. The representation of the confocal images into a 3D space of parameters acts as an interesting dimension reduction technique allowing classification algorithms to be implemented in quasi real-time. The GGD statistical model is adjusted to the intensities of the RCM images at different depths, to identify the skin depths at which lentigo detection and characterization are the most significant. A quantitative analysis supported by an SVM classifier is conducted to evaluate the performance of the proposed characterization.  A second contribution of this work is a new estimation algorithm for the GGD parameters, based on a natural gradient approach \cite{Amari}. The main property of this algorithm is its fast convergence compared to other existing techniques, allowing big databases to be processed with reduced computational cost. This approach is also known as Fisher scoring \cite{halimi2013parameter}. It updates the parameters in a Riemannian space, resulting in a fast convergence to a local minimum of the cost function of interest \cite{pereyra2013exploiting}. The proposed model and estimation algorithm are validated using synthetic and real RCM images, resulting from a clinical study containing healthy and lentigo patients. The obtained results are very promising.

 The paper is organized as follows. Section  2 presents the proposed method for lentigo identification. Simulation results are presented and analyzed in Section 3. Conclusions and perspectives for future works are finally reported in Section 4.

\section{Proposed method}
\begin{figure}[h]
\begin{center}
\includegraphics[width=12cm, height=6cm]{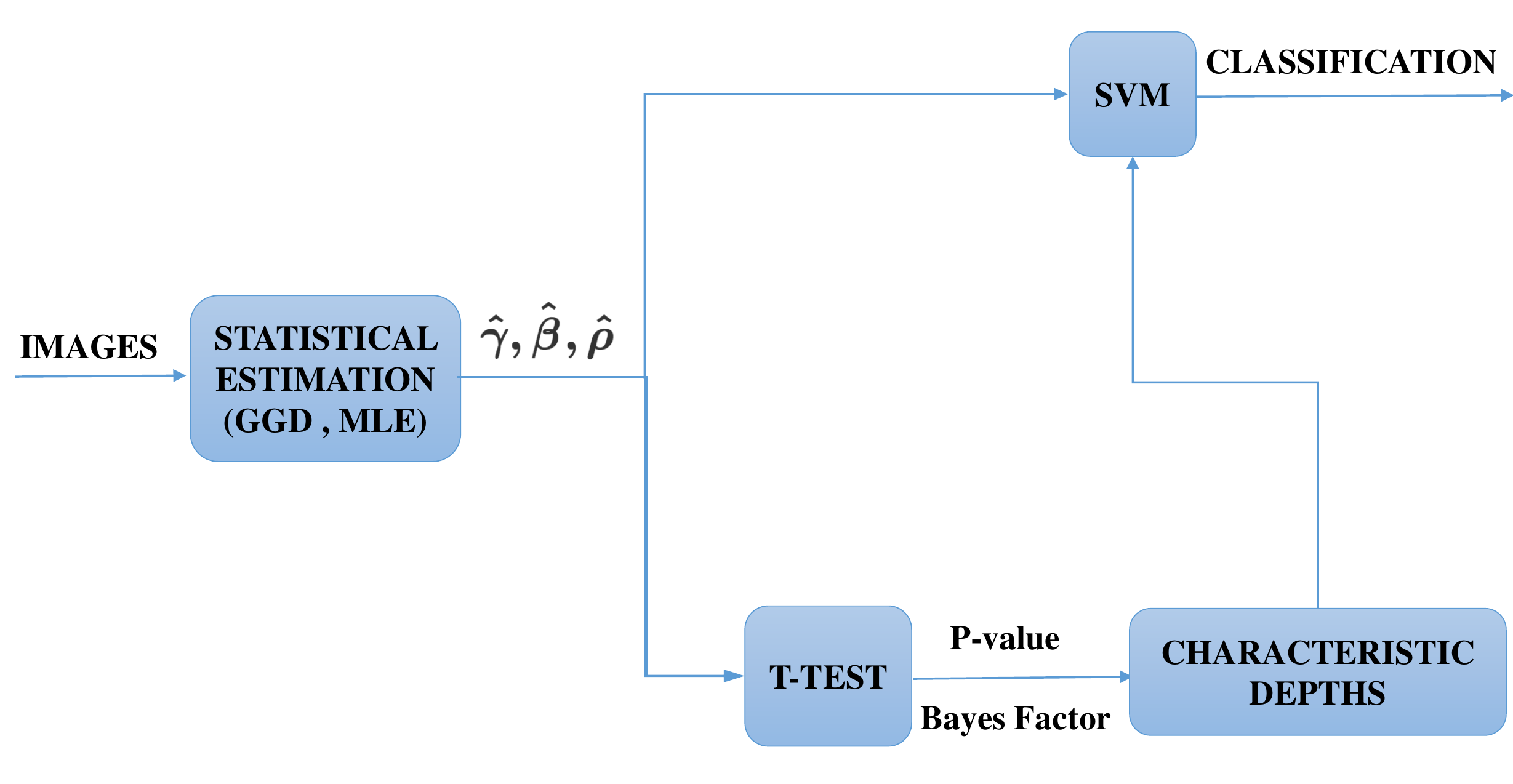}
\caption{\label{fig:method}Proposed classification method.}
\end{center}
\end{figure}

This section presents the proposed approach based on the use of the generalized gamma distribution to classify and characterize healthy and lentigo RCM images. It contains three steps that are summarized in Fig. \ref{fig:method} and described in the next sections.

\subsection{Statistical estimation}

\subsubsection{Generalized gamma distribution}
In this paper, we propose to use the statistical properties of the pixel intensities to detect the presence of lentigo in the RCM images. More precisely, we consider $L$ noise free images $\boldsymbol{x^l} = (x^l_1,\dots, x^l_N)$, containing $N$ pixels and we assume  that the distribution of these pixel intensities is a generalized gamma distribution \cite{stacy1,kotz2000continuous}. The GGD depends on the parameter vector $\boldsymbol{\theta_l}=\left(\gamma_l,\beta_l,\rho_l\right)$ and is defined as
\begin{equation}
f\left({x}^{l}_n; \boldsymbol{\theta_l} \right)=\begin{cases}\frac{1}{\beta_l^{\rho_l}\  \Gamma(\rho_l)} \ (x^l_n-\gamma_l)^{\rho_l-1}\exp\left(-\frac{x^l_n-\gamma_l}{\beta_l}\right), &  x^l_n>\gamma_l \ \ \ \beta_l,\rho_l>0 \\0 & \text{otherwise.} \end{cases}
\label{equa0}
\end{equation}

where $\Gamma(.)$ is the gamma function \cite{Abramowitz}. We will show in Section 3.2 that the density \eqref{equa0} is perfectly adapted to the distribution of the intensity values of an RCM image. The next section introduces a statistical estimation method based on the maximum likelihood principle which allows the parameter  $\boldsymbol{\theta}=(\boldsymbol{\theta}_1,\dots,\boldsymbol{\theta}_L)$ to be estimated from the samples $\boldsymbol{x}^{l}$. 

\subsubsection{Maximum likelihood approach}

The maximum likelihood (ML) estimation method consists of maximizing the likelihood of the observed samples with respect to the unknown model parameters \cite{Kay1993}. 
Assuming that the observations $x_n^l$ are independent, the likelihood function of the sample $\boldsymbol{x}^l= (x_1^l, ...,x_N^l)$ is defined as
\begin{equation}\small
f\left(\boldsymbol{x}^l ; \boldsymbol{\theta_l}\right)=\prod^{N}_{n=1}\; f\left({x}_n^{l}; \boldsymbol{\theta_l} \right)
\label{equaL}
\end{equation}

The log-likelihood function is then defined as:

\begin{equation}
\mathcal L(\boldsymbol{\theta_l})=\log\left[f\left(\boldsymbol{x}^l ; \boldsymbol{\theta_l}\right)\right]=-N\rho_l \ \log(\beta_l)-N\ \log\left[\Gamma(\rho_l)\right]+(\rho_l-1)\sum_{n=1}^N \log(x^l_n-\gamma)-\sum_{i=1}^N \frac{(x^l_n-\gamma_l)}{\beta_l}.
\label{eq_3}
\end{equation}
 The partial derivatives of the log-likelihood with respect to $\gamma_l$, $\beta_l$ and  $\rho_l$ can be computed easily, leading to

\begin{equation}
\it \begin{cases}\frac{\partial \ \mathcal L(\boldsymbol{\theta_l})}{\partial \gamma_l}=-(\rho_l-1) \  \sum_{n=1}^N (x^l_n-\gamma_l)^{-1}+\frac{N}{\beta_l} \\\frac{\partial \ \mathcal L(\boldsymbol{\theta_l})}{\partial \beta_l}=\frac{-N \rho_l}{\beta_l}+\frac{1}{\beta_l^{2} } \sum_{n=1}^N (x^l_n-\gamma_l) \\ \frac{\partial \ \mathcal L(\boldsymbol{\theta_l})}{\partial \rho_l}=-N \Psi(\rho_l)-N\ \log(\beta_l)+\sum_{n=1}^N \log(x^l_n-\gamma_l)   \end{cases}
\label{eq_4}
\end{equation}

where $\Psi\left(z\right)=\Gamma^{'}\left(z\right)/\Gamma\left(z\right)$ is the digamma function \cite{Abramowitz}.

 Maximizing the log-likelihood in \eqref{eq_3} can be obtained by making the derivatives in \eqref{eq_4}  equal to 0. Akin to Jonkman and al. \cite{Jonkman},  a good estimate can be obtained  using a gradient descent algorithm as follows: \ :
	
\begin{equation}
\boldsymbol{\theta_l^{t+1}}=\boldsymbol{\theta_l^t}+\lambda \ . A(\boldsymbol{\theta_l^t}) . \ \nabla \ \mathcal L(\boldsymbol{\theta_l^t})
\end{equation}

where \ $\nabla$ is the gradient operator,\; $\lambda$ is the step-size and $A$ a preconditioning matrix that depends on $\boldsymbol{\theta_l^t}$ (Hessian).\\

In our case, we have used the natural gradient descent in order to estimate the parameter $\boldsymbol{\theta_l}$. Unlike Newton's method, natural gradient adaptation does not assume a locally-quadratic cost function and for maximum likelihood estimation tasks, it is asymptotically Fisher-efficient \cite{Amari}.

	\begin{equation}
\boldsymbol{\theta_l^{t+1}}=\boldsymbol{\theta_l^t}+\frac{\lambda}{\parallel F^{-1}(\boldsymbol{\theta_l^t}).\ \nabla \mathcal L(\boldsymbol{\theta_l^t})\parallel}. \ F^{-1}(\boldsymbol{\theta_l^t}) \ \nabla \  \mathcal L(\boldsymbol{\theta_l^t})
\end{equation}
where $\nabla \mathcal L(\boldsymbol{\theta_l^t})$ is the gradient defined in \eqref{eq_4} and $F(\boldsymbol{\theta_l^t})$ is the Fisher matrix defined as

\begin{equation}
\small
F(\boldsymbol{\theta_l^t})= -  \begin{pmatrix}
E\left(\frac{\partial^2 \mathcal L(\boldsymbol{\theta_l^t})}{\partial \gamma_l^2}\right)&E\left(\frac{\partial^2 \mathcal L(\boldsymbol{\theta_l^t})}{\partial \gamma_l \  \partial\beta_l}\right) &E\left(\frac{\partial^2 \mathcal L(\boldsymbol{\theta_l^t})}{\partial \gamma_l \  \partial\rho_l}\right)\\
E\left(\frac{\partial^2 \mathcal L(\boldsymbol{\theta_l^t})}{\partial \beta_l \  \partial\gamma_l}\right)&E\left(\frac{\partial^2 \mathcal L(\boldsymbol{\theta_l^t})}{\partial \beta_l^2}\right) &E\left(\frac{\partial^2 \mathcal L(\boldsymbol{\theta_l^t})}{\partial \beta_l \  \partial\rho_l}\right)\\
E\left(\frac{\partial^2 \mathcal L(\boldsymbol{\theta_l^t})}{\partial \rho_l \  \partial\gamma_l}\right)&E\left(\frac{\partial^2 \mathcal L(\boldsymbol{\theta_l^t})}{\partial \rho_l \  \partial\beta_l}\right) &E\left(\frac{\partial^2 \mathcal L(\boldsymbol{\theta_l^t})}{\partial \rho_l^2}\right)
\end{pmatrix}.
\end{equation}

Straightforward  computation leads to:  \\

\begin{equation}
F(\boldsymbol{\theta_l^t})=  {N  \begin{pmatrix}
\frac{1}{\beta_l^2(\rho_l-2)}&\frac{1}{\beta_l^2} &\frac{1}{\beta_l(\rho_l-1)}\\
\frac{1}{\beta_l^2}&\frac{\rho_l}{\beta_l^2} &\frac{1}{\beta_l}\\
\frac{1}{\beta_l(\rho_l-1)}&\frac{1}{\beta_l} &\Psi'(\rho_l)
\end{pmatrix}}
\end{equation}\\
where {$\Psi'$} denotes the trigamma function \cite{Abramowitz}. The interest of using this natural gradient method for generalized gamma distributions will be clarified in the next section. To our knowledge, it is the first time that a natural gradient method is applied to generalized gamma distributions. The resulting estimates of parameters $\gamma, \beta$ and $\rho$ will be used to characterize the lentigo in RCM images, and to classify healthy and lentigo patients. Simulation results confirming these properties are presented in the next section.

\vspace{-0.2cm}
\subsection{Characterization using the parametric T-test}
\label{sec:test}
{We propose to apply a parametric T-test to the parameters $\gamma_l$, $\beta_l$ and $\rho_l$ to assess the statistical significance of these parameters abilities to separate healthy and lentigo patients.

We consider a corpus composed of images from $n_1$ healthy patients and $n_2$ lentigo patients, as annotated by an expert.
Let $\boldsymbol{\theta}_{H}= \left\{\boldsymbol{\theta}_{H}^1, \dots, \boldsymbol{\theta}_{H}^{n_1}\right\},$ denote the set of parameters estimated for healthy patients and $\boldsymbol{\theta}_{S}= \left\{\boldsymbol{\theta}_{S}^1, \dots, \boldsymbol{\theta}_{S}^{n_2}\right\}, $ for the lentigo patients at a given depth.  The MLE is known to be asymptotically Gaussian
and asymptotically efficient \cite{hurlin2013,scholz1985maximum}, and it is assumed that these parameters are independent and follow normal distributions.

A two-sample T-test \cite{cressie1986,wendorf2004,rakotomalala2013} can then be applied to compare the means $\mu_{H}$ and $\mu_{S}$
\begin{small}
\begin{align}
\label{eq:ho}
H_{0}  &:  \mu_{H}=\mu_{S}, \hspace{1cm}\\
\label{eq:hl}
H_{1}  &:  \mu_{H} \neq \mu_{S}. \hspace{1cm}
\end{align}
\end{small}
The variances of the distributions $\sigma^2_{H}$ and $\sigma^2_{S}$ being unknown and not equal, we propose to apply the test as follows.
Let $\bar{\boldsymbol{\theta}}_{H}$ and $\bar{\boldsymbol{\theta}}_{S}$ denote the empirical means of $\boldsymbol{\theta}_{H}$ and $\boldsymbol{\theta}_{S}$. Denoting as $s^2_{\boldsymbol{H}}$ and $s^2_{\boldsymbol{S}}$ the unbiased estimators of the variances for healthy and lentigo patients.\\
The statistics of the T-test associated with \eqref{eq:ho} and \eqref{eq:hl} is then
\begin{small}
\begin{equation}
 T = \frac{\bar{\boldsymbol{\theta}}_{H} - \bar{\boldsymbol{\theta}}_{S}}{\sqrt{\frac{s^2_{\boldsymbol{H}}}{n_1} + \frac{s^2_{\boldsymbol{S}}}{n_2}}}.
\end{equation}
\end{small}
that is distributed according to a Student T distribution with $\nu$ degrees of freedom, where $\nu$ is defined as
\begin{small}
\begin{equation}
\nu=    \frac{\left[\frac{s^2_{H}}{n_1}+\frac{s^2_{S}}{n_2}\right]^2}{\frac{s^4_{H}}{n^2_1(n_1-1)}+\frac{s^4_{S}}{n^2_2(n_2-1)}}.
\end{equation}
\end{small}
The hypothesis $H_{0}$ is rejected if $\left|T_{\nu}\right| > T_{PFA}$. In this study, we chose a probability of false alarm $PFA=0.05$ corresponding to a threshold $T_{PFA}=2.02$ for $n_1=18$ and $n_2=27$.} To assess the statistical significance, the p-value of each test has been also calculated, and the following decision rules have been applied
\begin{itemize}
	\item When $p$ value $>  0.10$    \  $\rightarrow$ the observed difference is ``not significant''
	\item When $p$ value $\in \left[0.05, 0.10\right]$ \  $\rightarrow$ the observed difference is ``marginally significant''
	\item When $p$ value $\in [0.01, 0.05[$ \  $\rightarrow$ the observed difference is ``significant''
	\item When $p$ value $<  0.01$ \  $\rightarrow$ the observed difference is ``highly significant''.
\end{itemize}
Given the recent debate on the p-value and the reproducibility of scientific results, a method has been developed in \cite{johnson2013} to establish a correspondence between classical significance tests, such as the one designed here, with Bayesian tests.  This method allows one to relate the size of the classical hypothesis tests with evidence thresholds in Bayesian tests. Following this work and assuming equal variances, we calculated the Bayes factor ($BF$) given by
\begin{small}
\begin{equation}
BF = \left( \frac{\nu + T}{\nu + \left(T - \sqrt{(\nu \alpha^*)}\right)^2} \right)^{(n_1+n_2)/2}
\end{equation}
\end{small}
where the hypothesis $H_{0}$ is rejected when $BF>\sqrt{\nu \alpha^*}$ with $\alpha^* = \alpha^{2/(n_1+n_2-1)}-1$ and $\alpha = [(T_{PFA}^2/\nu) +1]^{(n_1+n_2)/2}$. This BF provided the same evidence as the considered p-values as will be shown in the experimental part.

\vspace{-0.2cm}
\section{Experiments}
\subsection{Performance of the proposed estimation algorithm}
\subsubsection{Synthetic data}
\begin{figure}[htbp]
\centering
\includegraphics[width=16cm]{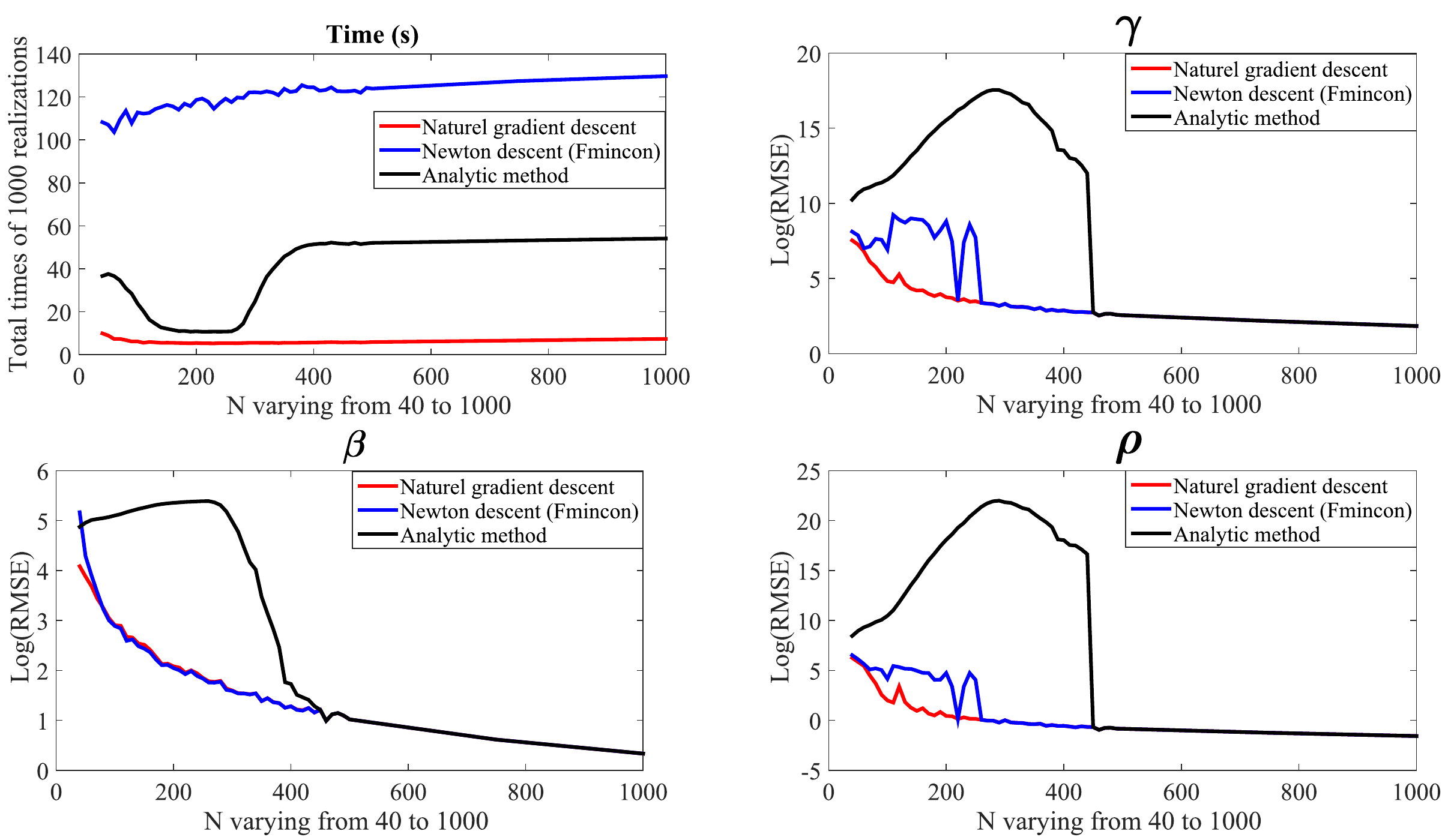}
\caption{\small Evolution of the RMSE and the total time of the three methods for N varying from 40 to 1000.}
\label{fig:time_rmse}
\end{figure}

\begin{figure}[htbp]
\centering
\includegraphics[width=16cm]{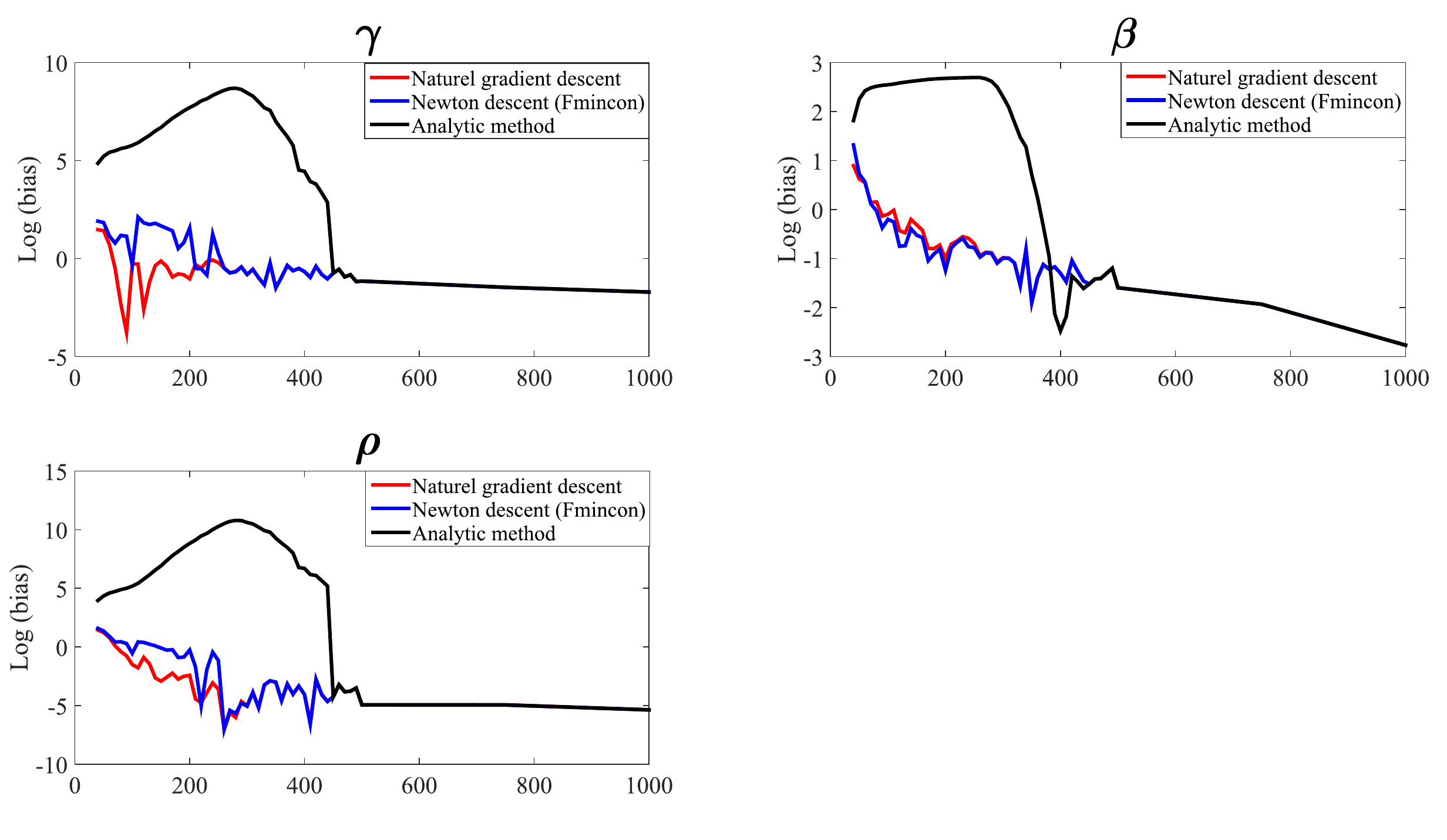}
\caption{\small Evolution of the bias of the three methods for N varying from 40 to 1000.}
\label{fig:bias}
\end{figure}

This section evaluates the performance of the proposed estimation algorithm on
synthetic data. The hardware used here was a consumer-quality PC with an Intel(R) Core(TM) i7-4860HQ CPU 2.4 GHz processor, 32 GB RAM, and an Nvidia GeForce GTX 980m graphics card, running 64-bits Windows 10. The algorithm was applied using the software MATLAB R2014b. Two experiments were conducted, the first one is based on the generation of samples varying from $N$= 40 to 1000 using \eqref{equa0} with the following fixed parameters $\boldsymbol{\theta}$=($\gamma=2$, $\beta=15$, $\rho=4$). The
second experiment is done by considering one parameter variable and the other two constants for two case $N$=100 and $N$=10000. The algorithm was run M=1000 realizations for the first experiment and M=100 for the second one. The estimated parameters $\boldsymbol{\hat{\theta}}$ was then used to calculate the bias, the variance and the root mean square 
error (RMSE) which are defined as follow : 

\begin{equation}
\boldsymbol{\hat{\theta}} =  { \begin{pmatrix}
\hat{\gamma} \\
\hat{\beta} \\ 
\hat{\rho}
\end{pmatrix}}
\end{equation}  
\begin{equation}
\text{Bias}=E\left(\boldsymbol{{\theta}}-\boldsymbol{\hat{\theta}}\right)= \left(\boldsymbol{{\theta}}-\frac{\sum_{i=1}^M{\boldsymbol{\hat{\theta}}({i})}}{M}\right)
\end{equation}

\begin{equation}
\text{Variance}=E\left[\boldsymbol{\hat{\theta}}^{2}\right]-\left[E\left(\boldsymbol{\hat{\theta}}\right)\right]^{2}
\end{equation}

\begin{equation}
\text{RMSE}=\text{Bias}^{2}+\text{Variance}=\frac{\sum^{M}_{i=1}\left({\boldsymbol{\hat{\theta}}({i})}-\boldsymbol{{\theta}}\right)^{2}}{M}
\end{equation}
where $\boldsymbol{\hat{\theta}}({i})$ denotes the estimated parameters for the $i$th realization.\\
Our method was compared to two other methods, the first one was proposed in  \cite{Jonkman} and applied a Newton gradient descent using the hessian as a pre-conditioned matrix. The second one (denoted by analytical method) used the solution of the null derivatives (see \eqref{eq_4}) to have a new formulation of the parameters $\beta$ and $\rho$ only depending on the $\gamma$ parameter. The resulting problem depends only on the $\gamma$ parameter as follows
\begin{equation}
-N \Psi(\rho_l)-N\ \log(\beta_l)+\sum_{n=1}^N \log(x^l_n-\gamma_l)=0
\label{eq_mthd3}
\end{equation}
with  
 \begin{equation*}
\beta_l=\frac{\left[\sum_{n=1}^N \left(x_n^l-\gamma_l\right) \ \sum_{n=1}^N \left(\frac{1}{x_n^l-\gamma_l}\right)\right]-N^2}{N \ \sum_{n=1}^N \left(\frac{1}{x_n^l-\gamma_l}\right)}
\end{equation*}
 \begin{equation*}
\rho_l=\frac{\sum_{n=1}^N \left(x_n^l-\gamma_l\right) \ \sum_{n=1}^N \left(\frac{1}{x_n^l-\gamma_l}\right)}{\left[\sum_{n=1}^N \left(x_n^l-\gamma_l\right) \ \sum_{n=1}^N \left(\frac{1}{x_n^l-\gamma_l}\right)\right]-N^2}.
\end{equation*}
This problem is solved using a Newton gradient descent algorithm. The bias and the RMSEs of the parameter estimates are displayed in log scale (to improve readability) in Figs. \ref{fig:time_rmse} and \ref{fig:bias} with the associated running times. Note that the three methods were initialized with the same estimator based on the ``pseudo method of moments'' (see \cite{Jonkman} for details). The proposed method based on a natural gradient recursion provides smaller RMSEs and bias for a small number of samples, i.e, for $N \in \{40, ...,300 \}$. This proves that our method is more efficient than the other methods for a small number of samples. The natural gradient descent also provides a faster convergence compared to the other methods with a significant reduction in computational cost for any sample size. This result is interesting since it allows big databases of RCM images to be processed more easily. These results highlight the good performance of the proposed strategy for the estimation of GGD parameters.

The second experiment is done by considering $N$=100 and $N$=10000 samples when varying one of the parameters ($\gamma$, $\beta$, $\rho$)  and fixing the others.

We considered 10 values for the variable parameter and achieved M=100 realisations to estimate $\hat{\theta}$.

\begin{figure}[htbp]
\centering
\includegraphics[width=17cm]{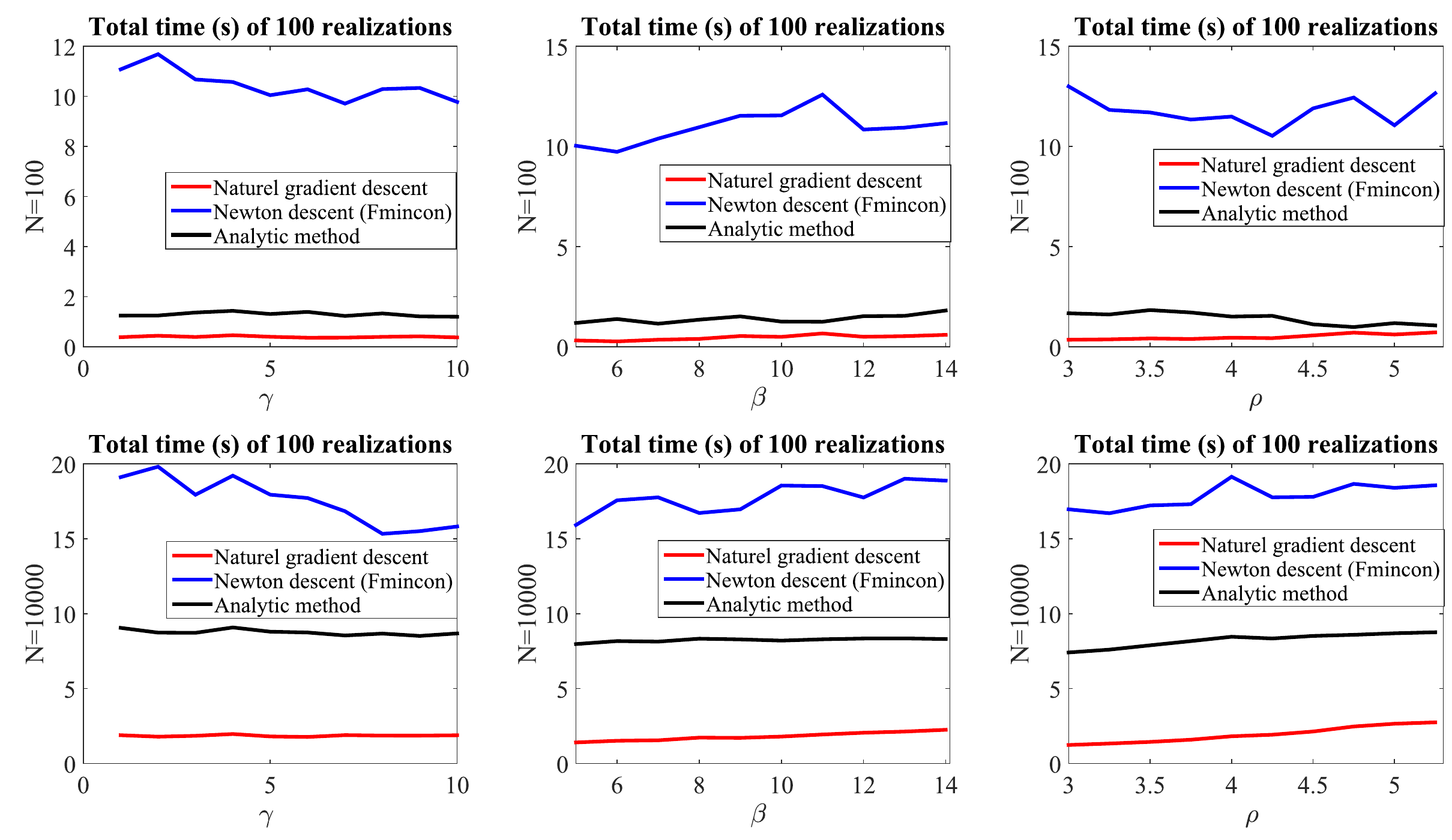}
\caption{\small Evolution of  the total time of the three methods for N=10000 and N=100 when varying the parameters ($\gamma$, $\beta$, $\rho$) .}
\label{fig:time_n10000_n100}
\end{figure}

\begin{figure}[htbp]
\centering
\includegraphics[width=16cm]{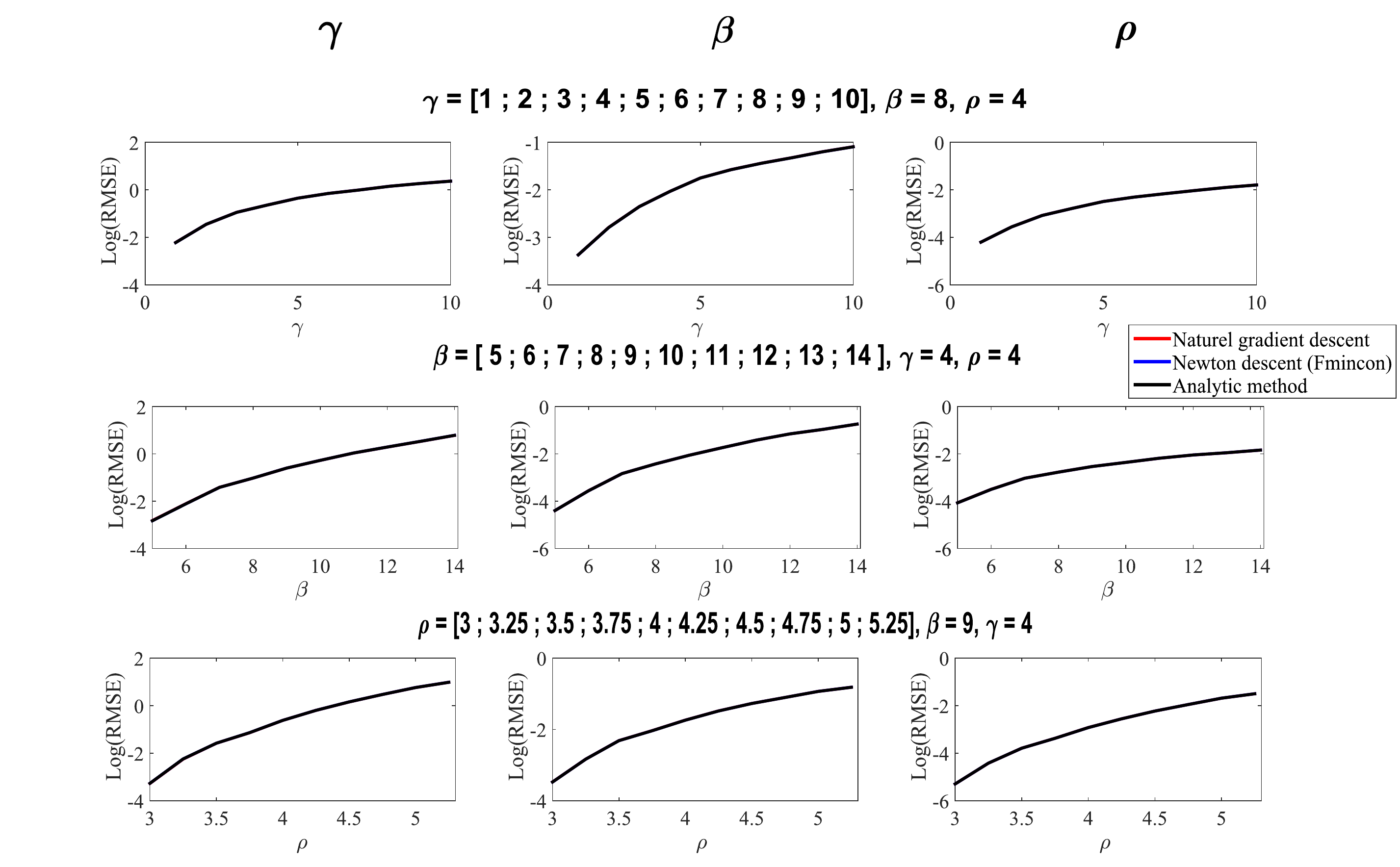}
\caption{\small Evolution of the RMSE of the three methods for N=10000 when varying the parameters ($\gamma$, $\beta$, $\rho$).}
\label{fig:rmse_n10000}
\end{figure}

\begin{figure}[htbp]
\centering
\includegraphics[width=16cm]{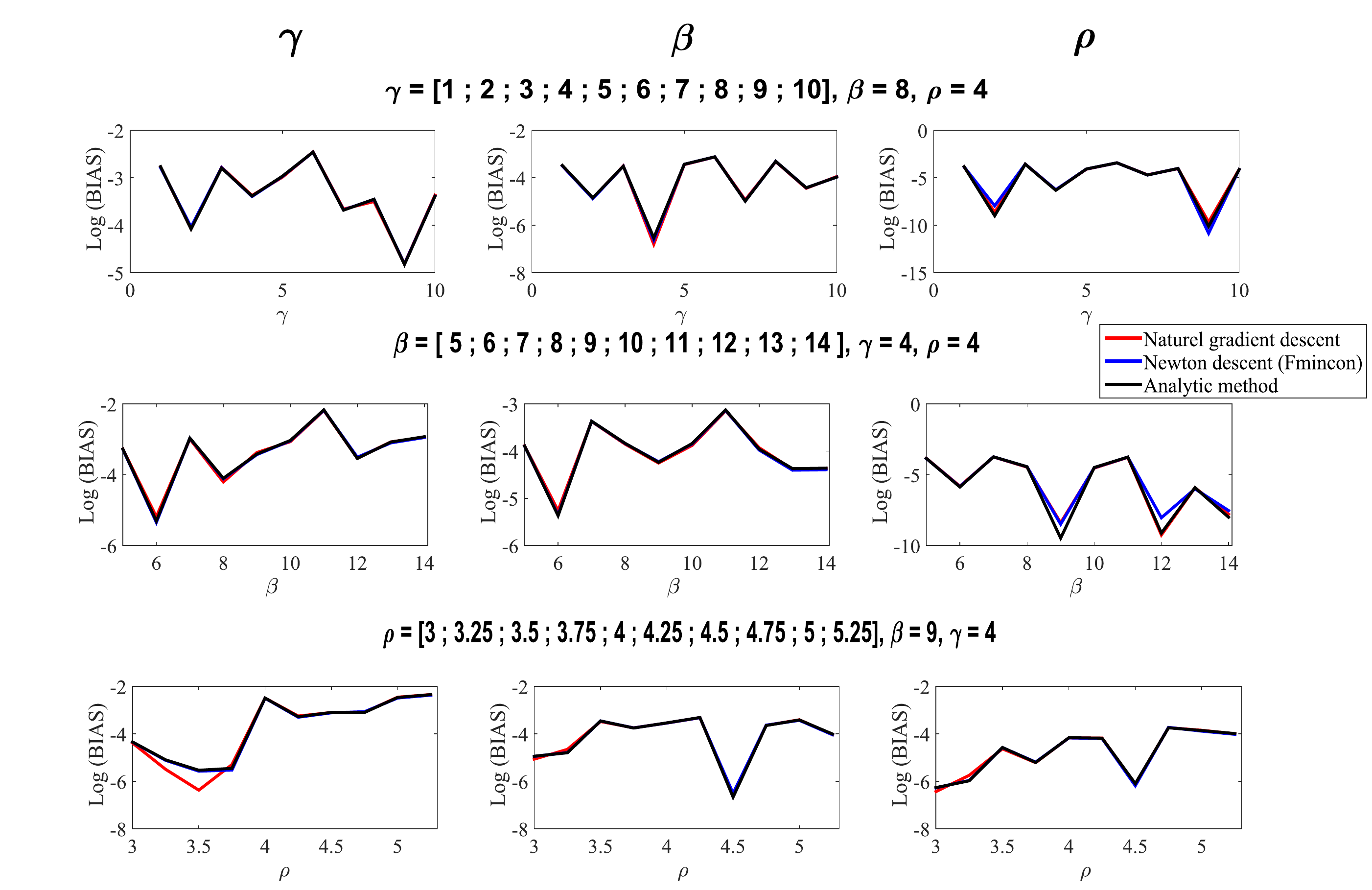}
\caption{\small Evolution of the bias of the three methods for N=10000 when varying the parameters ($\gamma$, $\beta$, $\rho$).}
\label{fig:bias_n10000}
\end{figure}

\begin{figure}[htbp]
\centering
\includegraphics[width=16cm]{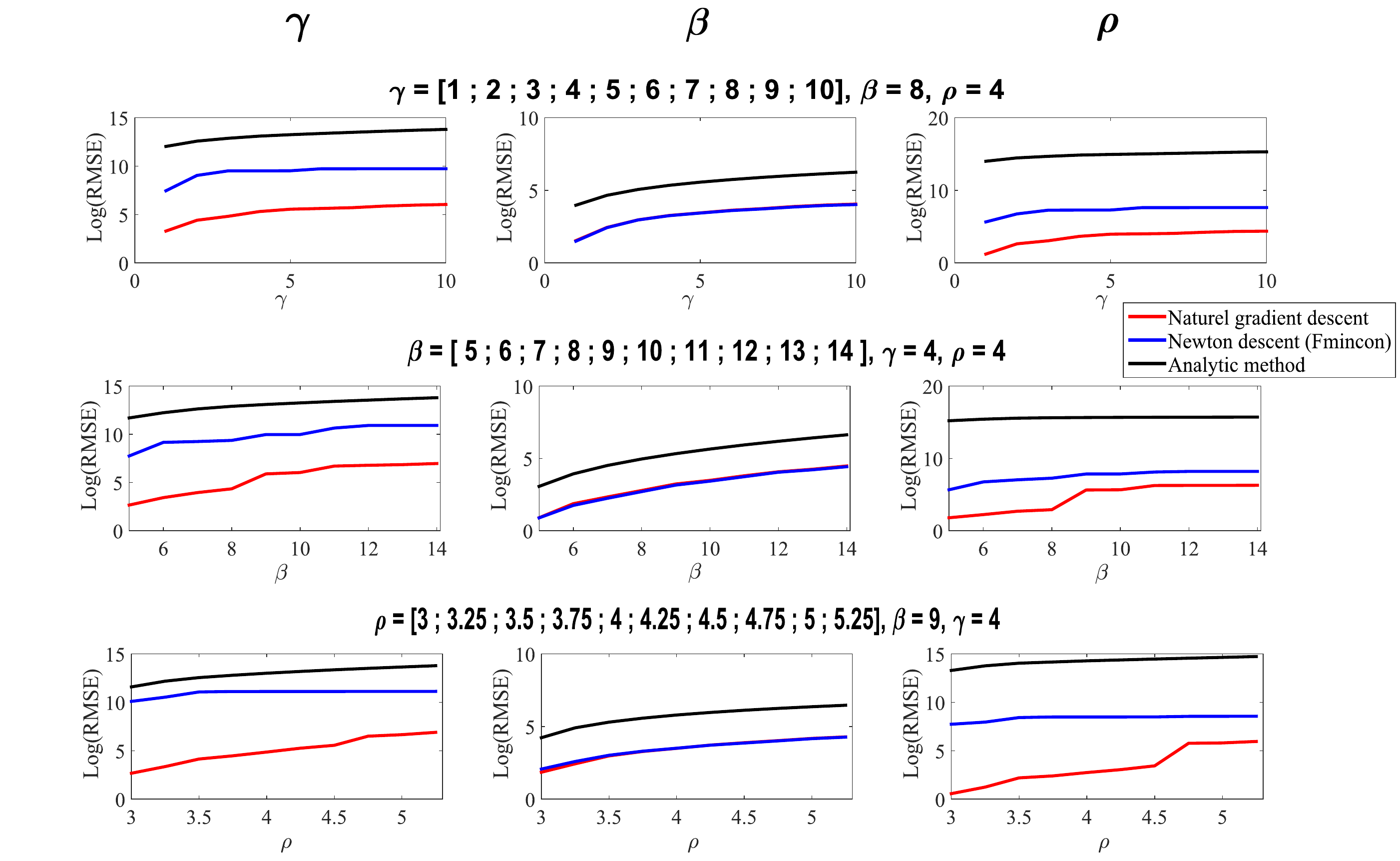}
\caption{\small Evolution of the RMSE of the three methods for N=100 when varying the parameters ($\gamma$, $\beta$, $\rho$).}
\label{fig:rmse_n10000}
\end{figure}

\begin{figure}[htbp]
\centering
\includegraphics[width=16cm]{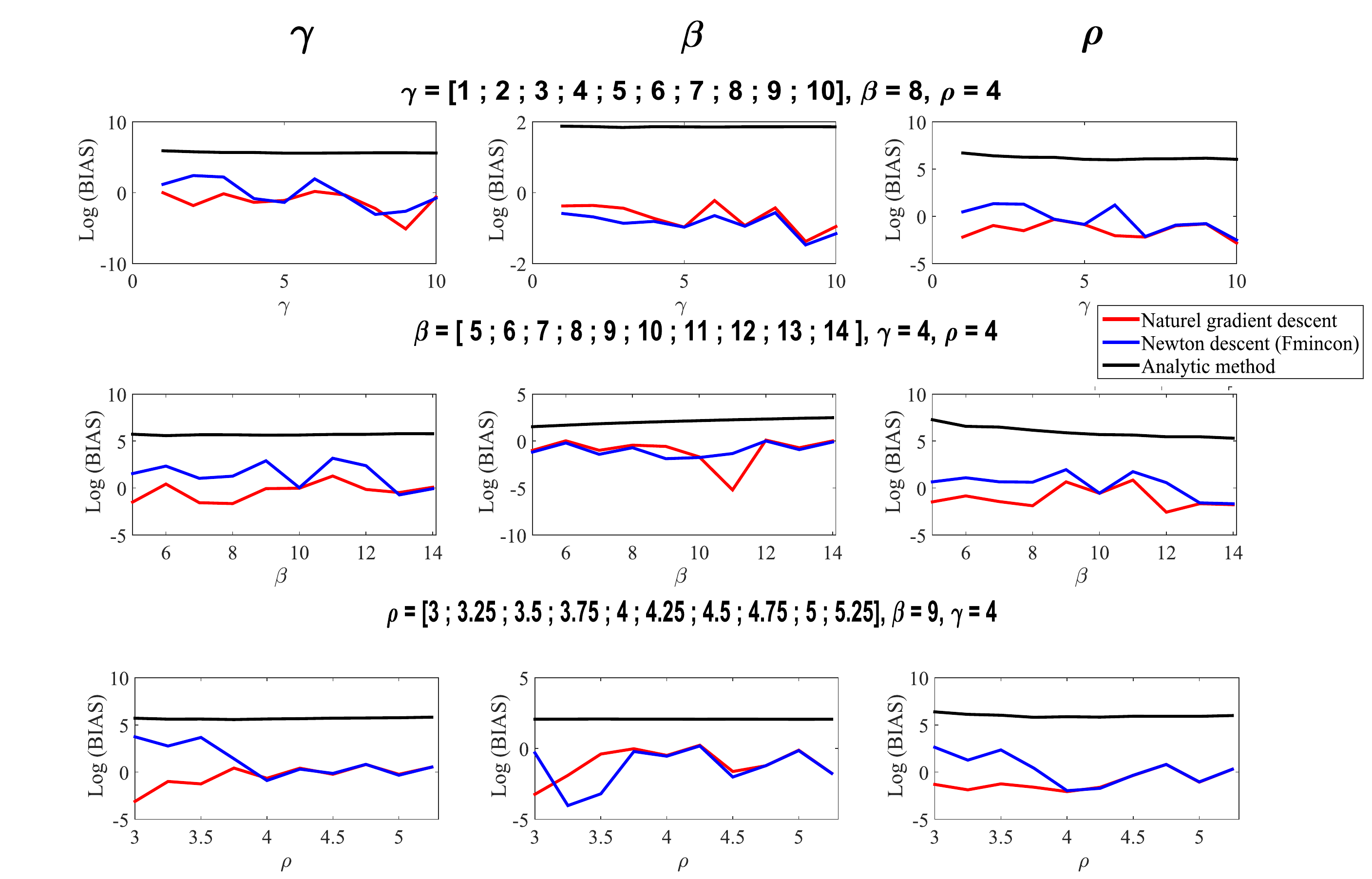}
\caption{\small Evolution of the bias of the three methods for N=100 when varying the parameters ($\gamma$, $\beta$, $\rho$).}
\label{fig:bias_n10000}
\end{figure}

The results obtained for the second experiment in Figs. 4 to 8 confirm the conclusions of the first experiment. Our method has a faster convergence compared to the other methods and this regardless of the number of samples and is more efficient than the other methods for the case of small number of samples ($N$=100).

\newpage
\vspace{3cm}
\subsubsection{Clinical RCM data}
\begin{figure*}[htbp]
\centering
\includegraphics[width=16cm, height=8cm]{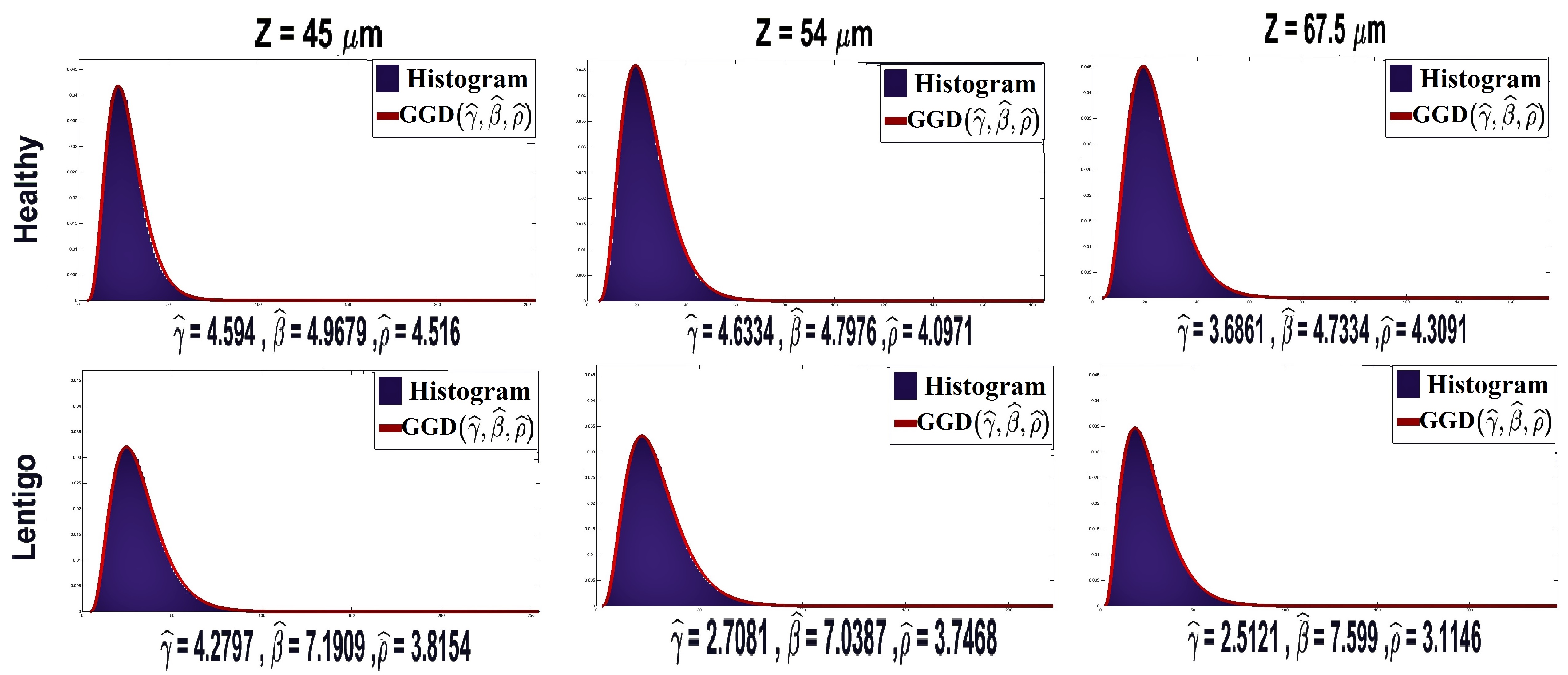}
\caption{\small Histograms of the  intensities and the corresponding estimated GGD distributions. The figure shows data from two arbitrary healthy and lentigo patients ($\#6$ and $\#38$ respectively) at three representative depths (one depth per column).}
\label{fig:fit}
\end{figure*}

This section is devoted to the validation of the proposed  algorithm when applied to real RCM images, which were  performed with apparatus Vivascope $1500$. The in-vivo images were acquired from the stratum corneum, the epidermis layer, the dermis-epidermis junction (DEJ) and the upper papillary dermis. Each RCM image shows a $500 \times 500 \mu m$ field of view with $1000 \times 1000$ pixels.  A set of $L=45$ women aged $60$ years and over were recruited.  All the volunteers gave their informed consent for examination of skin by RCM. According to the clinical evaluation performed by a physician, volunteers were divided into two groups.  The first group was formed by $27$ women with at least $3$ lentigines on the back of the hand whereas $18$ women without lentigo  constituted the control group. Two acquisitions were performed on each volunteer for the $25$ depths. Images were taken on lentigo lesions for volunteers of the first group and on healthy skin on the back of the hand for the control group. An examination of each acquisition was performed in order to locate the stratum corneum and the DEJ precisely  in each image. Consequently, our database contained $L$ = 45 patients. For each patient, we retained two stacks of 25 RCM images, giving a total of 2250 images. Fig. \ref{fig:fit} compares the histograms of the intensities of the RCM images with the estimated GGD distributions at $3$ representative depths. This figure concerns two arbitrary healthy and lentigo patients, namely patients $\#6$ and $\#38$ respectively. It illustrates the goodness of fit of the generalized gamma distribution for the intensities of the RCM images for both healthy and lentigo images. A difference in the scale and the shape of the distributions can be observed between healthy and lentigo patients, as illustrated by the differences in the corresponding parameters $\beta$ and $\rho$. These differences are at the basis of the proposed characterization method as illustrated in the next experiments. 

\begin{figure}[htbp]
\centering
\includegraphics[width=14cm]{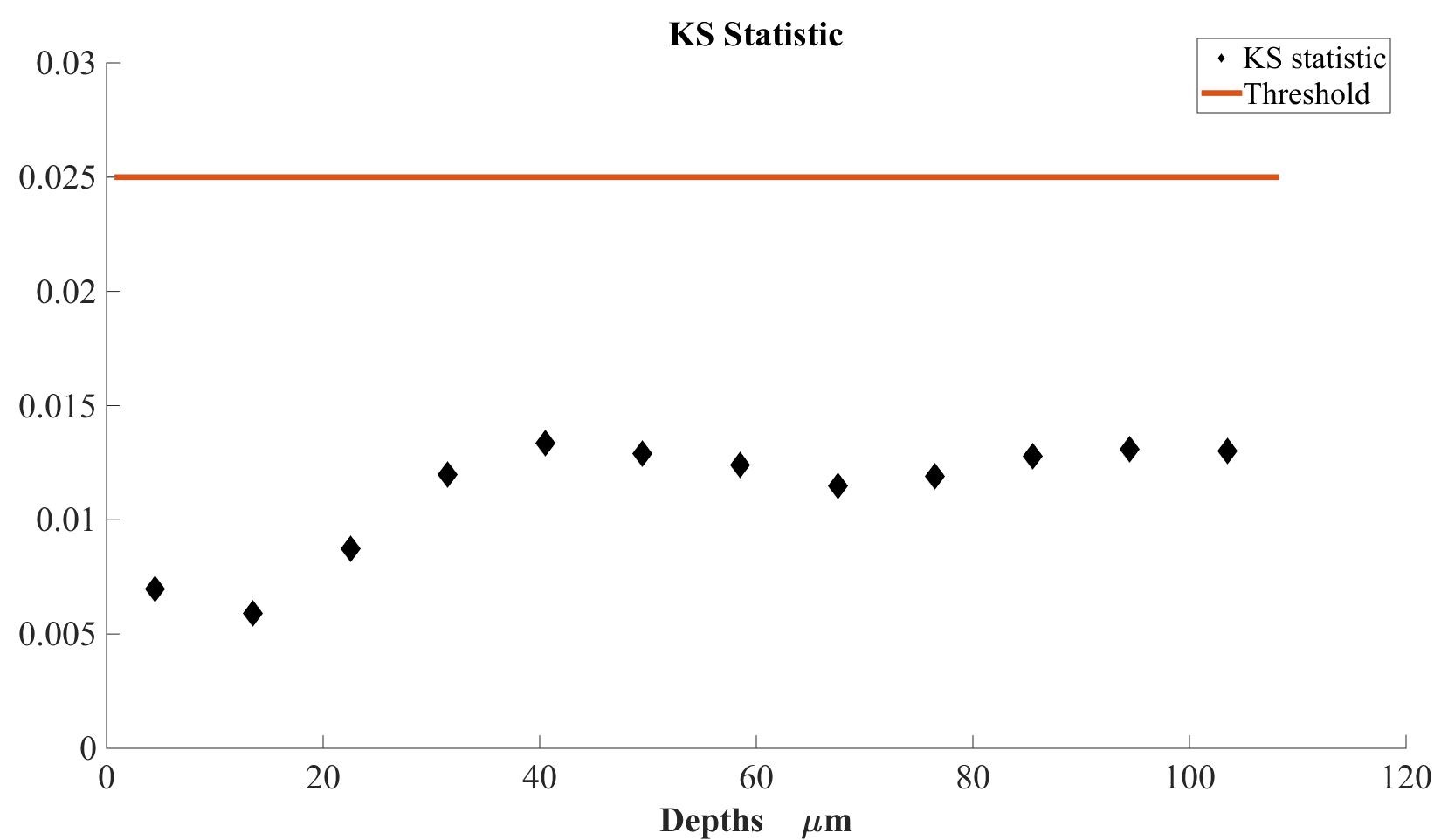}
\caption{\small Assessment of the GGD fit to the pixels. Mean KS statistic for the whole population for all depths. Scores are excellent for all configurations.}
\label{fig:ks}
\end{figure}

Fig. \eqref{fig:ks} shows the quantitative assessment of the fit using the the Kolmogorov-Smirnov (KS) test. The mean KS statistic score of the whole population ($45$ patients) has been calculated at each depth. One notices the excellent scores with KS values very close to zero.

\subsection{Identification of characteristic depths}
\begin{figure}[htbp]
\centering
\includegraphics[width=16cm, height=7cm]{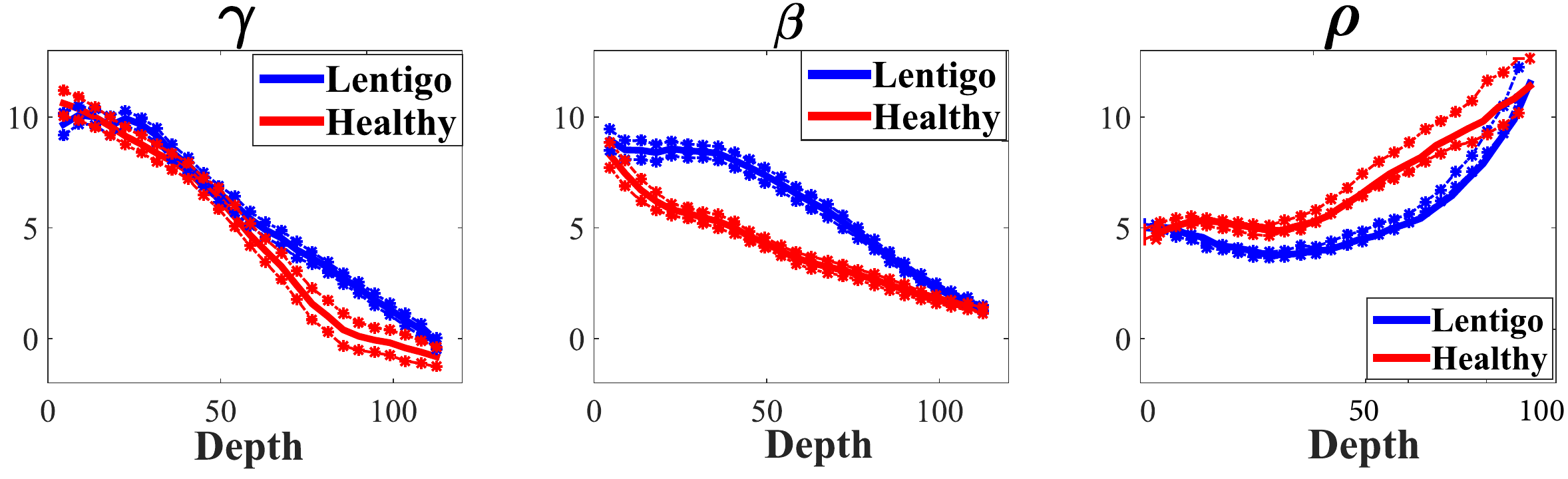}
\caption{\small Evolution of the average parameters $\hat{\gamma}$, $\hat{\beta}$ and $\hat{\rho}$ throughout the depth. Values of $\gamma$ are too similar for healthy and lentigo patients and cannot be used for discrimination. The parameter $\beta$ and $\rho$ shows significant difference for depths between $30{\mu}m$ and $76{\mu}m$, with maximal difference at around $50{\mu}m$. Our conclusion is that the parameters $\beta$ and $\rho$ can discriminate healthy and lentigo skin tissues.}
\label{fig:tparams}
\end{figure}

\begin{figure*}[htbp]
\centering
\includegraphics[width=13cm, height=10cm]{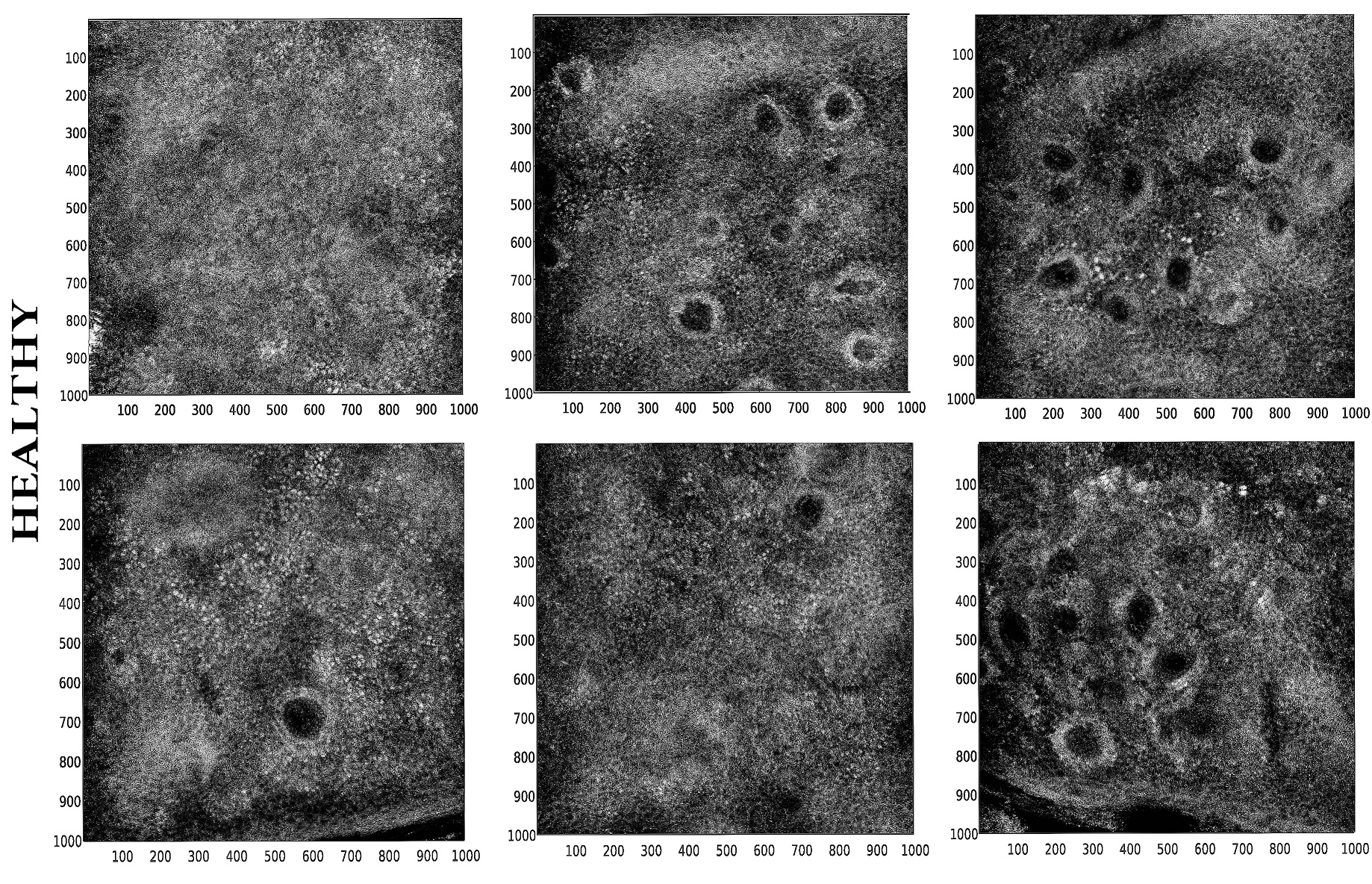}\\
\includegraphics[width=13cm, height=10cm]{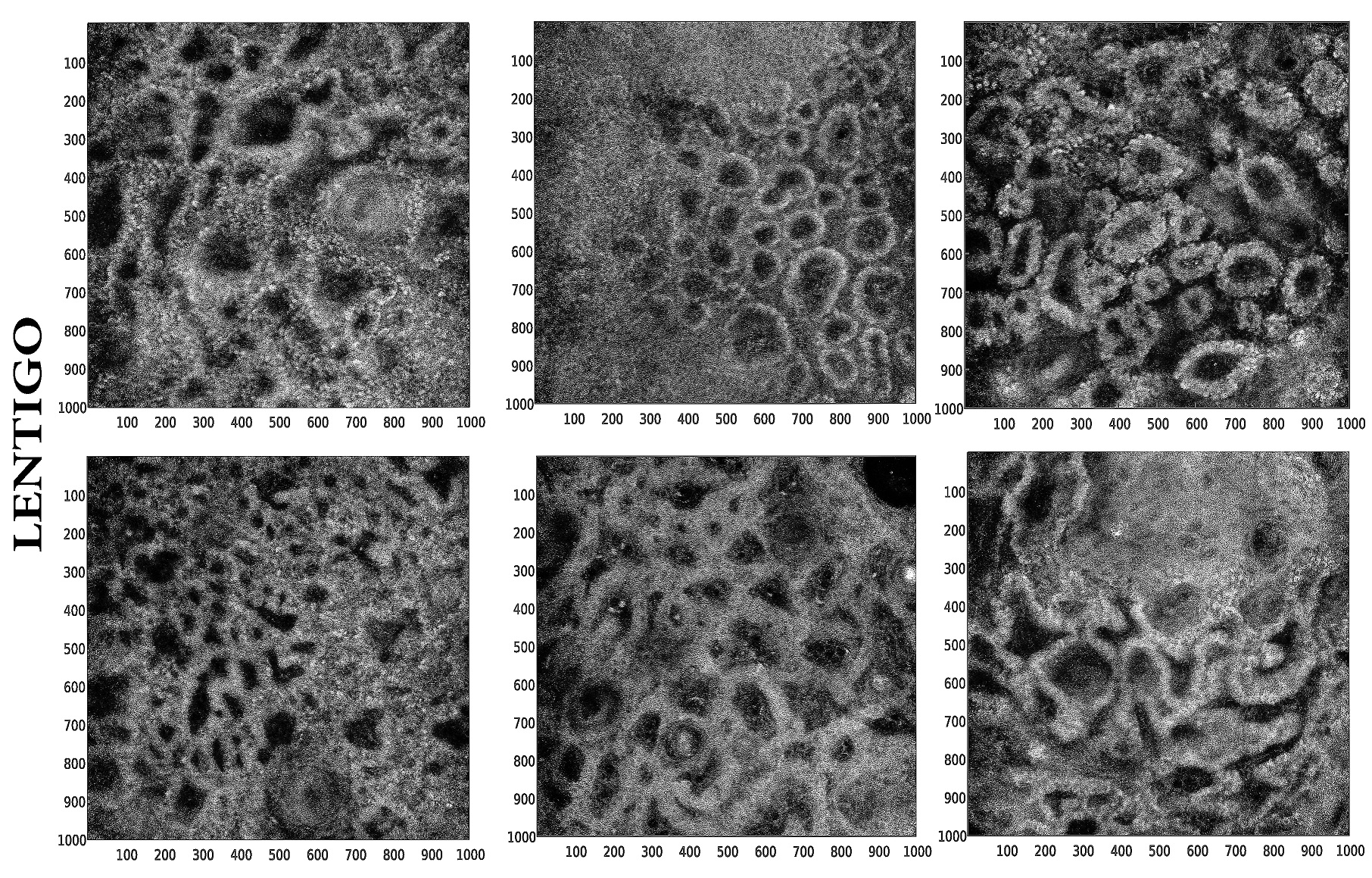}
\caption{\small Images from healthy (patient $\#1$, $\#2$, $\#3$, $\#4$, $\#5$, $\#6$) and lentigo (patient $\#31$, $\#33$, $\#37$, $\#38$, $\#40$, $\#44$) patients at the DEJ depth (54 $\mu$m). One can observe more textured images in the presence of lentigo.}
\label{fig:images}
\end{figure*}

The GGD distributions were fitted to each of the $2250$ images at each depth. Having acquired two stacks of $25$ images for each patient, one of the two stacks was selected randomly for the analysis.  The average parameters $\overline{\boldsymbol{\theta}}_{Healthy}$ and $\overline{\boldsymbol{\theta}}_{Lentigo}$ were then calculated at each depth for healthy and lentigo patients, respectively. To account for variability, the process of selecting one stack for each patient was repeated $300$ times and average curves and standard deviations were calculated. Results are shown in Fig. \ref{fig:tparams}, which clearly shows that  $\gamma$ does not characterize healthy and lentigo patients. Conversely, $\beta,\rho$ allows the discrimination of healthy and lentigo images for depths between $30{\mu}m$ and $76{\mu}m$, with maximal difference at around $50{\mu}m$.  Fig. \ref{fig:images} shows two sets of images associated with six healthy $(\#1,\#2,\#3,\#4,\#5,\#6)$ and  six lentigo $(\#31,\#33,\#37,\#38,\#40,\#44)$ patients. One can observe more textured images in the presence of lentigo at the DEJ depth (corresponding to $54{\mu}m$), as expected.

\subsection{Statistical significance with T-Test}
\label{sec:stats_test}
\begin{figure}[htbp]
\centering
\includegraphics[width=16cm]{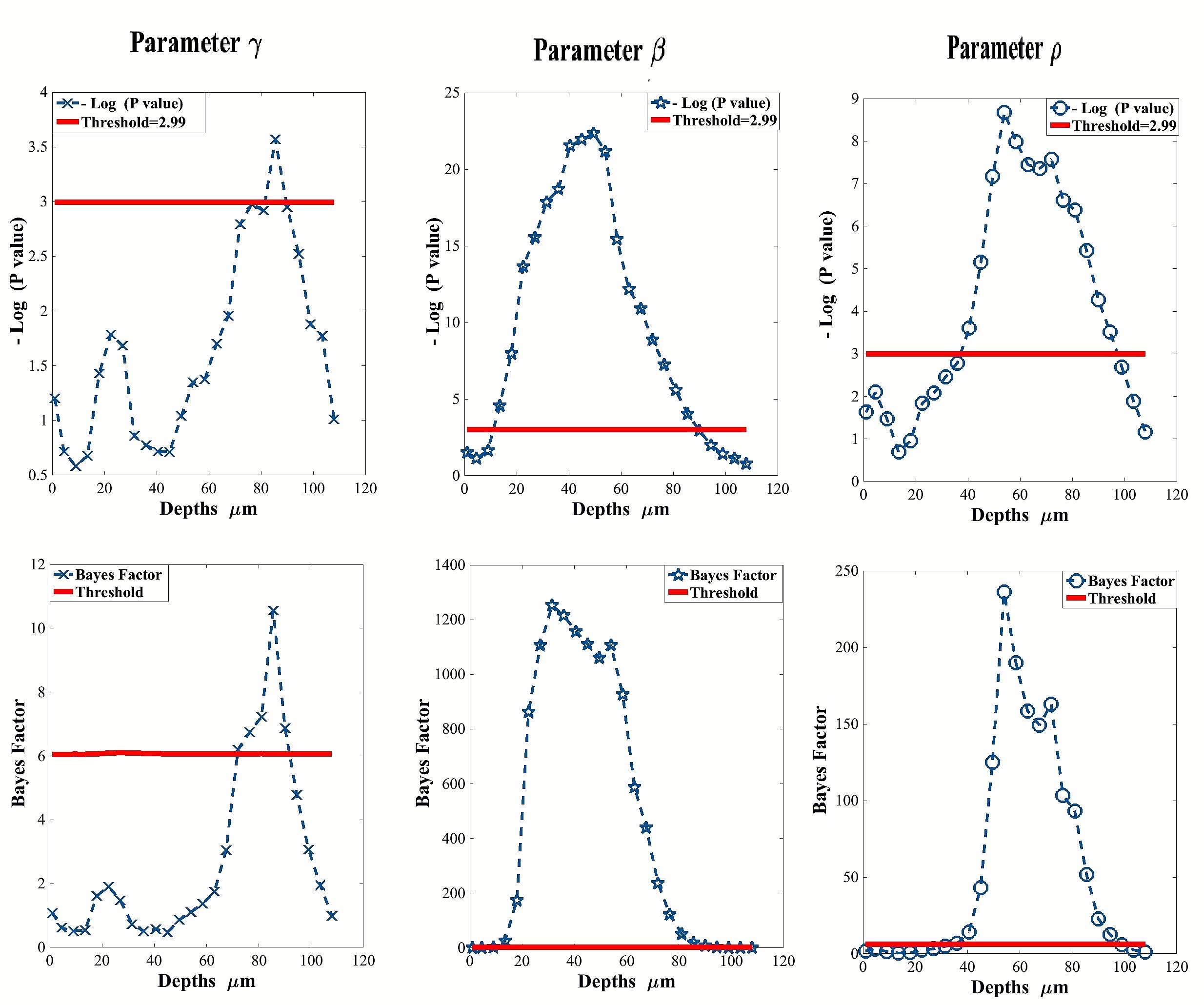}
\caption{\label{fig:ttestbeta}P-value (in $-log$ scale) and Bayes factor (BF) of the T test for $\gamma$, $\beta$ and $\rho$. The weak score shows that $\gamma$ is not  discriminant between healthy and lentigo images. Strong scores can be seen for depths between $30{\mu}m$ and $76{\mu}m$ and highest scores are obtained with depths around $50{\mu}m$ for the parameters $\beta$ and $\rho$. This confirms that $\beta$ and $\rho$ are good discriminating functions that can be used to separate healthy and lentigo images at these depths.}
\end{figure}

\begin{figure}[htbp]
\centering
\includegraphics[width=15cm]{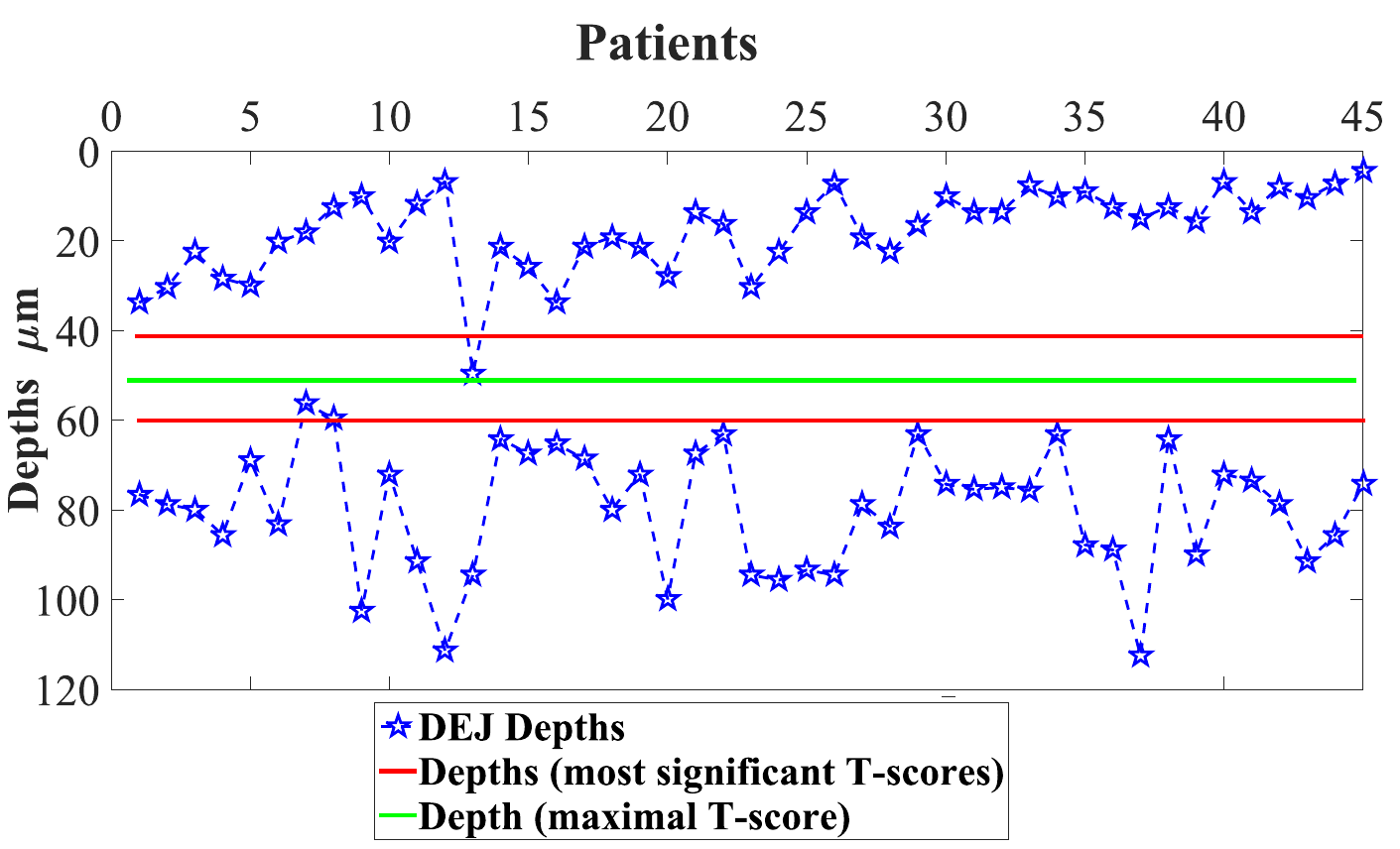}
\caption{\label{fig:dejdepths}Characteristics depths (found to be between 40um and 60um according to the T-test) and DEJ depths associated with the 45 patients. }
\end{figure}

The parametric test described in Section \ref{sec:test} has been applied to the parameters $\gamma$, $\beta$ and $\rho$ at each depth to assess the significance of the results. Fig. \ref{fig:ttestbeta} shows the p-values and Bayes factors associated with the two-sample T-Test conducted respectively with $\gamma$, $\beta$ and $\rho$, at different depths.
The p-value has been represented in $-\log$ scale for readability. This figure shows weak values for $\gamma$ both for the p-value and the Bayes factor confirming that $\gamma$ cannot discriminate healthy and lentigo images. On the other hand it shows high scores for both indicators for a range of depths for the parameters $\beta$ and $\rho$. Table \ref{tab:highbeta} presents depths that give $T_{\beta}$ and $T_{\rho}$ higher than the threshold $T_{PFA}=2.02$, hence confirming the hypothesis that the parameters $\beta$ and $\rho$ can be used to discriminate healthy and lentigo patients. This table also shows the depths that provide p-values lower than the probability of false alarm $PFA = 0.05$ and their corresponding Bayes factor. According to our decision rules, the results are \emph{highly significant} for depths between $40\mu m$ and $60\mu m$, with the highest score at $49.5\mu m$ for $\beta$ and $54\mu m$ for $\rho$. These results are in good agreement with the quantitative differences shown in Fig. \ref{fig:tparams}. These results confirm that $\beta$ and $\rho$ give a good test statistics for discriminating lentigo and healthy skin, especially at depths around $50{\mu}m$. As mentioned in the introduction, lentigines are mainly characterized in RCM by the disorganization of the dermoepidermal junction (DEJ). Coherently, the parameter $\beta$ is very discriminant at depths close to $50 \mu m$, which corresponds to the average location of the range of depths that represent the DEJ (annotated by the dermatologists) as shown in Fig. \ref{fig:dejdepths}.

\begin{table*}[]
\centering
\renewcommand{\arraystretch}{1.2}
\setlength{\tabcolsep}{0.2cm}
\begin{tabular}{cc|c|c|c|c|}
\cline{3-6}
                                                                             &                     & \multicolumn{2}{c|}{$\beta$}          & \multicolumn{2}{c|}{$\rho$} \\ \cline{3-6} 
                                                                             &                     & min               & max               & min          & max          \\ \hline
\multicolumn{1}{|c|}{\multirow{4}{*}{$T_{\text{scores}}$ \textgreater 2.02}} & depths              & 13.5              & 86                & 40           & 94           \\
\multicolumn{1}{|c|}{}                                                       & $T_{\text{scores}}$ & 2.6               & 2.46              & 2.33         & 2.27         \\
\multicolumn{1}{|c|}{}                                                       & p-value             & 0.0106            & 0.018             & 0.0273       & 0.0302       \\
\multicolumn{1}{|c|}{}                                                       & BF                  & 26.5              & 18                & 14.03        & 12.64        \\ \hline
\multicolumn{1}{|c|}{\multirow{4}{*}{Significant $T_{\text{scores}}$}}       & depths              & 38.25             & 60.75             & 42.75        & 65.25        \\
\multicolumn{1}{|c|}{}                                                       & $T_{\text{scores}}$ & 7.90              & 6.11              & 2.8          & 3.8          \\
\multicolumn{1}{|c|}{}                                                       & p\_value            & $5.6 . 10^{-5}$   & $5.31 . 10^{-4}$  & 0.0101       & 0.0067       \\
\multicolumn{1}{|c|}{}                                                       & BF                  & 1177              & 862               & 40           & 151          \\ \hline
\multicolumn{1}{|c|}{\multirow{4}{*}{Maximal $T_{\text{scores}}$}}           & depths              & \multicolumn{2}{c|}{49.5}             & \multicolumn{2}{c|}{54}     \\
\multicolumn{1}{|c|}{}                                                       & $T_{\text{scores}}$ & \multicolumn{2}{c|}{8.54}             & \multicolumn{2}{c|}{4.21}   \\
\multicolumn{1}{|c|}{}                                                       & p-value             & \multicolumn{2}{c|}{$2.78 . 10^{-5}$} & \multicolumn{2}{c|}{0.0018} \\
\multicolumn{1}{|c|}{}                                                       & BF                  & \multicolumn{2}{c|}{1060}             & \multicolumn{2}{c|}{236}    \\ \hline
\end{tabular}
\vspace{0.2cm}
\caption{Depths where $T^{(\beta)} \text{and} \ T^{(\rho)} > T_{0.05}=2.02$; corresponding p-value and Bayes factor ($BF$) are shown. The first row gives intervals of depths (min depth to max depth) where T-scores are significant. The second row shows the depths giving highest T-scores (maximal T-score $\mp$ $10\%$). The third row shows the depths corresponding to the maximal T-score. P-values and Bayes factors corresponding to each depth are shown below.}
\label{tab:highbeta}
\end{table*}

\subsection{Performance of the SVM classifier}
\begin{figure*}[htbp]
\centering
    \includegraphics[scale=0.4]{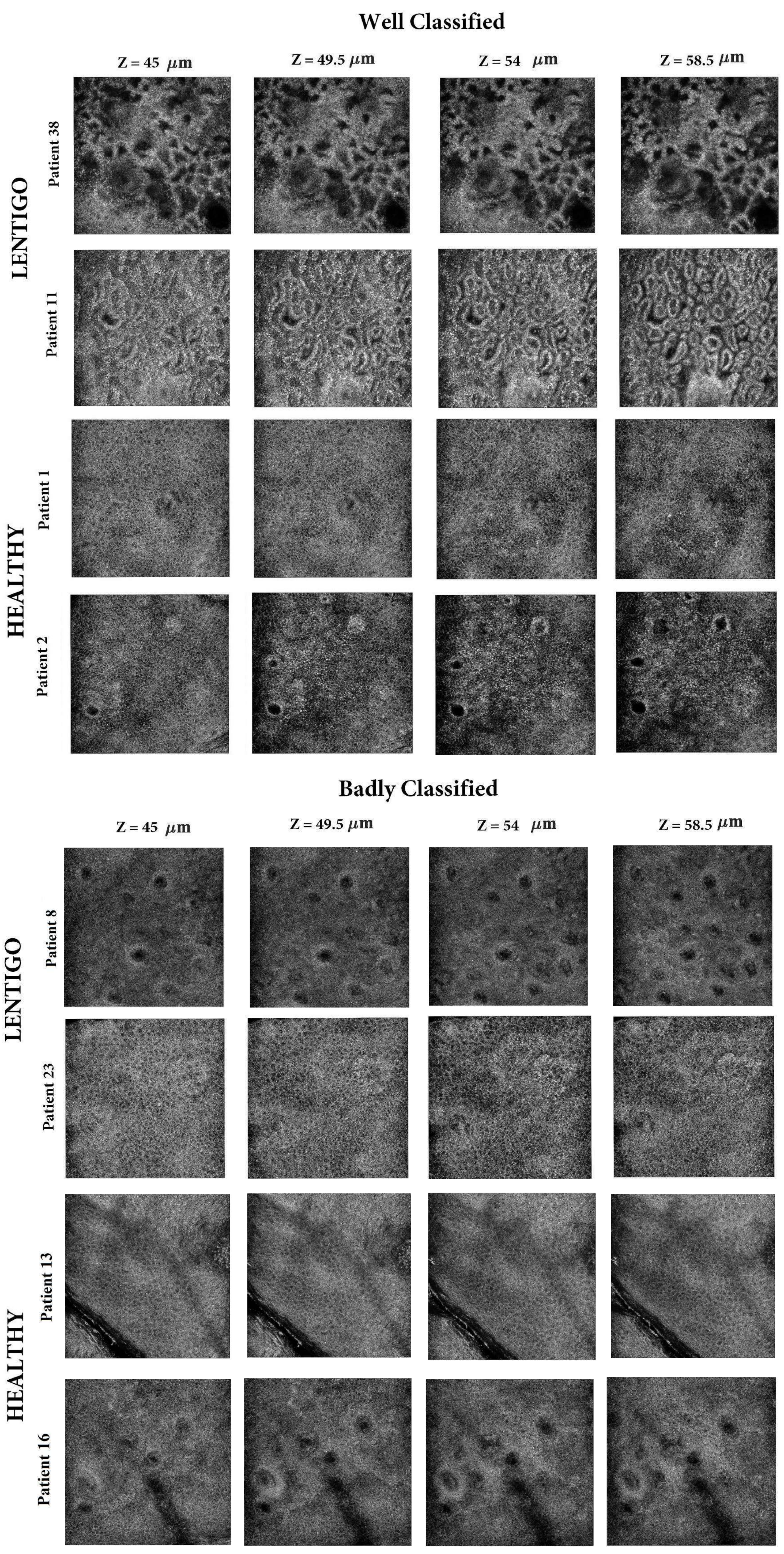}\\
\caption{\small Examples of RCM images of lentigo and healthy patients classified by the SVM classifier.}
\label{fig:svm images}
\end{figure*}

The section \ref{sec:stats_test} provided an estimate for the depths characterising
healthy from lentigo patients. Based on these depths we considered a simple SVM classification algorithm
to confirm their validity. The GGD parameters associated with these characteristic depths $(40\mu m \  \text{to} \ 60\mu m)$  were then used to classify the patients into 2 classes referred to as ``lentigo'' and ``healthy''. The leave-one-out method was used to compute the different probability of errors. This method uses $L-1$ images for training (where $L$ is the number of patients in the database) and the remaining image for testing. This operation was run $M=1000$ times. For each experiment, we considered only images from one acquisition out of the available two (for each patient). The obtained $M$ results were used to calculate the average confusion matrix shown in Table \ref{tab:svm} and to evaluate the average indicators (Sensitivity, Specificity, Precision and Accuracy). These indicators are defined as Sensitivity = TP/(TP+FN), Specificity = TN/(FP+TN), Precision = TP/(TP+FP), Accuracy = (TP+TN)/(TP+FN+FP+TN), where TP, TN, FP and FN are the numbers of true positives, true negatives, false positives and false negatives. This table allows us to assess the classification performance for these characteristic depths.  The results show that the classification of healthy and lesion tissues has an accuracy equal to $84.4$ \% for $\beta$ and equal to $82.2$ \% for $\rho$, thus we recommend to use the parameter $\beta$ for this classification.  Fig. \ref{fig:svm images} shows examples of classified RCM images using the proposed methodology. To assess the significance of our results, our algorithm was then compared to the method presented in \cite{koller2011vivo}. This method consists of extracting from each RCM image a set of $39$ parameters (further technical details are available in \cite{wiltgen2008automatic}) and applying to these features a classification procedure based on a classification and regression tree (CART). Note that the CART algorithm was tested on the real RCM images using a leave one out procedure. As shown in Table \ref{tab:Performance_on_real_cart}, the accuracy obtained with the CART algorithm is $80 \%$, i.e., it is slightly smaller than the one obtained with the proposed method and leads to two additional mis-classified patients. Moreover, the estimated GGD parameters can be used for the characterization of RCM images, which is not possible with CART.

\begin{table}[]
\centering
\normalsize 
\renewcommand{\arraystretch}{1.5}
\setlength{\tabcolsep}{0.1cm}
\begin{tabular}{c|c|c|c|c|c|c|}
\cline{2-7}
                                       & \multicolumn{3}{c|}{$\beta$}                                                                                        & \multicolumn{3}{c|}{$\rho$}                                                                                          \\ \hline
\multicolumn{1}{|c|}{Confusion matrix} & $\hat{L}$              & $\hat{H}$              & \begin{tabular}[c]{@{}c@{}}Sensitivity\\ Specificity\end{tabular} & $\hat{L}$               & $\hat{H}$              & \begin{tabular}[c]{@{}c@{}}Sensitivity\\ Specificity\end{tabular} \\ \hline
\multicolumn{1}{|c|}{Lentigo}          & 22                     & 5                      & 81.4  \%                                            & 21                      & 6                      & 77.7   \%                                           \\ \hline
\multicolumn{1}{|c|}{Healthy}          & 2                      & 16                     & 88.8  \%                                            & 2                       & 16                     & 88.8  \%                                            \\ \hline
\multicolumn{1}{|c|}{Precision}        & 91.6  \% & 76.1  \% &                                                                   & 87.5   \% & 72.7  \% &                                                                   \\ \hline
\multicolumn{1}{|c|}{Accuracy}        & \multicolumn{3}{c|}{84.4  \%}                                                                         & \multicolumn{3}{c|}{82.2  \%}                                                                          \\ \hline
\end{tabular}
\vspace{0.3cm}
\caption{Confusion matrices of SVM classifiers based on $\beta$ and $\rho$. One notices the good accuracy for the two parameters especially for $\beta$, where the 5 misclassified patients are consistently $\{ \#8, \#17, \#23, \#25, \#33 \}$.}
\label{tab:svm}
\end{table}

\begin{table}[]
\centering
\caption{Classification performance on real data (45 patients) using the CART method.}
\label{tab:Performance_on_real_cart}
{\renewcommand{\arraystretch}{1.5}
\begin{tabular}{|c|c|c|c}
\hline
\textbf{Confusion matrix} &$\widehat{\textbf{L}}$ & $\widehat{\textbf{H}}$ & \multicolumn{1}{c|}{\textbf{\begin{tabular}[c]{@{}c@{}}Sensitivity\\ Specificity\end{tabular}}} \\ \hline
\textbf{Lentigo}          & \textbf{24}             & \textbf{3} & \multicolumn{1}{c|}{\textbf{88.8 $\boldsymbol\%$}}                                                                  \\ \hline
\textbf{Healthy}          & \textbf{5}                 & \textbf{13} & \multicolumn{1}{c|}{\textbf{72.2 $\boldsymbol\%$}}                                                                   \\ \hline
\textbf{Precision}        & \textbf{82.7 $\boldsymbol\%$}           & \textbf{81.2 $\boldsymbol\%$ }        & \textbf{}                                                                                       \\ \cline{1-3}
\textbf{Accuracy}         & \multicolumn{2}{c|}{\textbf{82.2 $\boldsymbol\%$}}          & \textbf{}                                                                                       \\ \cline{1-3}
\end{tabular}}
\end{table}

\section{Conclusions}
\vspace{-0.3cm}
This paper investigated the potential of using the statistical properties of the pixels intensities associated with reflectance confocal microscopy images to classify between healthy and lentigo patients. The proposed method computed the location, scale and shape parameters of a generalized gamma distribution for the RCM images. It was shown that the lentigo is mainly characterized in the DEJ depth. The parameters at this depth were then used to train an SVM classifier. The proposed classifier was tested on a database of $2250$ real images associated with $45$ patients. The obtained results showed that the scale parameter $\beta$ is well adapted to the characterization and classification of healthy and lesion tissues. 
\vspace{-0.2cm}

\bibliographystyle{IEEEbib}

\bibliography{biblio_all}

\begin{thebibliography}{10}

\bibitem{menge2016concordance}
T~D Menge, B~P Hibler, M~A Cordova, K~S Nehal, and A~M Rossi,
\newblock ``Concordance of handheld reflectance confocal microscopy (rcm) with
  histopathology in the diagnosis of lentigo maligna (lm): A prospective
  study,''
\newblock {\em J. Am. Acad. Dermatol.}, vol. 74, no. 6, pp. 1114--1120, 2016.

\bibitem{Nehal2008}
K.~S. Nehal, D.~Gareau, and M.~Rajadhyaksha,
\newblock ``Skin imaging with reflectance confocal microscopy,''
\newblock {\em Seminars in Cutaneous Medecine and Surgery, Elsevier}, vol. 27,
  pp. 37--43, 2008.

\bibitem{Hofmann2009}
R.~Hofmann-Wellenhof, E.M.T. Wurm, V.~Ahlgrimm-Siess, E.~Richtig, S.~Koller,
  J.~Smolle, and A.~Gerger,
\newblock ``Reflectance confocal microscopy-state-of-art and research
  overview,''
\newblock {\em Seminars in Cutaneous Medecine and Surgery, Elsevier}, vol. 28,
  pp. 172--179, 2009.

\bibitem{Alarconcar2014}
I.~Alarcon, C.~Carrera, J.~Palou, L.~Alos, J.~Malvehy, and S.~Puig,
\newblock ``Impact of in vivo reflectance confocal microscopy on the number
  needed to treat melanoma in doubtful lesions,''
\newblock {\em British journal of Dermatology.}, vol. 170, pp. 802--808, 2014.

\bibitem{Alarcon2014}
I~Alarcon, C~Carrera, L~Alos, J~Palou, J~Malvehy, and S~Puig,
\newblock ``In vivo reflectance confocal microscopy to monitor the response of
  lentigo maligna to imiquimod,''
\newblock {\em J. Am. Acad. Dermatol.}, vol. 71, pp. 49--55, 2014.

\bibitem{Guitera2014}
P.~Guitera, L.E. Haydu, S.W. Menzies, and al,
\newblock ``Surveillance for treatment failure of lentigo maligna with
  dermoscopy and in vivo confocal microscopy: new descriptors.,''
\newblock {\em British journal of Dermatology.}, vol. 170, pp. 1305--1312,
  2014.

\bibitem{champin2014vivo}
J.~Champin, J.L. Perrot, E.~Cinotti, B.~Labeille, C.~Douchet, G.~Parrau,
  F.~Cambazard, P.~Seguin, and T.~Alix,
\newblock ``In vivo reflectance confocal microscopy to optimize the spaghetti
  technique for defining surgical margins of lentigo maligna,''
\newblock {\em Dermatologic Surgery}, vol. 40, no. 3, pp. 247--256, 2014.

\bibitem{Hibler2015}
B~P Hibler, M~Cordova, R~J Wong, and A~M Rossi,
\newblock ``Intraoperative real-time reflectance confocal microscopy for
  guiding surgical margins of lentigo maligna melanoma,''
\newblock {\em Dermatologic Surgery.}, vol. 41, pp. 980--983, 2015.

\bibitem{Luck2005}
B.L. Luck, K.D. Carlson, A.C. Bovik, and R.R. Richards-Kortum,
\newblock ``An image model and segmentation algorithm for reflectance confocal
  images of in vivo cervical tissue,''
\newblock {\em IEEE Trans. Image Processing}, vol. 14, no. 9, pp. 1265--1276,
  2005.

\bibitem{harris2015pulse}
M.A. Harris, A.N. Van, B.H. Malik, J.M. Jabbour, and K.C. Maitland,
\newblock ``A pulse coupled neural network segmentation algorithm for
  reflectance confocal images of epithelial tissue,''
\newblock {\em PloS one}, vol. 10, no. 3, pp. e0122368, 2015.

\bibitem{kurugol2011semi}
S.~Kurugol, M.~Dy, J.G.and~Rajadhyaksha, K.~Gossage, J.~Weissmann, and D.H.
  Brooks,
\newblock ``Semi-automated algorithm for localization of dermal/epidermal
  junction in reflectance confocal microscopy images of human skin,''
\newblock in {\em SPIE BiOS}. International Society for Optics and Photonics,
  2011, pp. 79041A--79041A--10.

\bibitem{kurugol2012validation}
S.~Kurugol, M.~Rajadhyaksha, J.G. Dy, and D.H. Brooks,
\newblock ``Validation study of automated dermal/epidermal junction
  localization algorithm in reflectance confocal microscopy images of skin,''
\newblock in {\em SPIE BiOS}. International Society for Optics and Photonics,
  2012, pp. 820702--820702--11.

\bibitem{Somoza2014}
E.~Somoza, G.O. Cula, C.~Correa, and J.B. Hirsch,
\newblock {\em Automatic Localization of Skin Layers in Reflectance Confocal
  Microscopy}, pp. 141--150,
\newblock Springer International Publishing, Cham, 2014.

\bibitem{hames2015anatomical}
S~C Hames, M~Ardigo, H~P Soyer, A~P Bradley, and T~W Prow,
\newblock ``Anatomical skin segmentation in reflectance confocal microscopy
  with weak labels,''
\newblock in {\em Proc. Int. Conf. Dig. Image Comput. (dICTA'2015)}, Adelaide,
  AUS, 2015, pp. 1--8.

\bibitem{hames2016automated}
S~C Hames, M~Ardig{\`o}, H~P Soyer, A~P Bradley, and T~W Prow,
\newblock ``Automated segmentation of skin strata in reflectance confocal
  microscopy depth stacks,''
\newblock {\em PloS one}, vol. 11, no. 4, pp. e0153208, 2016.

\bibitem{kose2016machine}
K.~Kose, C.~Alessi-Fox, M.~Gill, J.G. Dy, D.H. Brooks, and M.~Rajadhyaksha,
\newblock ``A machine learning method for identifying morphological patterns in
  reflectance confocal microscopy mosaics of melanocytic skin lesions
  in-vivo,''
\newblock in {\em SPIE BiOS}. International Society for Optics and Photonics,
  2016, pp. 968908--968908--8.

\bibitem{koller2011vivo}
S.~Koller, M.~Wiltgen, V.~Ahlgrimm-Siess, W.~Weger, R.~Hofmann-Wellenhof,
  E.~Richtig, J.~Smolle, and A.~Gerger,
\newblock ``In vivo reflectance confocal microscopy: automated diagnostic image
  analysis of melanocytic skin tumours,''
\newblock {\em Journal of the European Academy of Dermatology and Venereology},
  vol. 25, no. 5, pp. 554--558, 2011.

\bibitem{raphael2013computational}
A.P. Raphael, T.A. Kelf, E.M.T. Wurm, A.V. Zvyagin, H.P. Soyer, and T.W. Prow,
\newblock ``Computational characterization of reflectance confocal microscopy
  features reveals potential for automated photoageing assessment,''
\newblock {\em Experimental dermatology}, vol. 22, no. 7, pp. 458--463, 2013.

\bibitem{Amari}
S.~Amari and S.C. Douglas,
\newblock ``Why natural gradient?,''
\newblock in {\em IEEE Int. Conf. Acoust., Speech, and Signal Processing
  (ICASSP)}, Seattle, WA, May 1998, vol.~2, pp. 1213--1216.

\bibitem{halimi2013parameter}
A~Halimi, C~Mailhes, J.-Y Tourneret, P~Thibaut, and F~Boy,
\newblock ``Parameter estimation for peaky altimetric waveforms,''
\newblock {\em IEEE Transactions on Geoscience and Remote Sensing}, vol. 51,
  no. 3, pp. 1568--1577, 2013.

\bibitem{pereyra2013exploiting}
M~Pereyra, H~Batatia, and S~McLaughlin,
\newblock ``Exploiting information geometry to improve the convergence
  properties of variational active contours,''
\newblock {\em IEEE Journal of Selected Topics in Signal Processing}, vol. 7,
  no. 4, pp. 700--707, 2013.

\bibitem{stacy1}
E.W. Stacy,
\newblock ``A generalization of the gamma distribution,''
\newblock {\em Annals of Mathematical Statistics}, vol. 33, no. 3, pp.
  1187--1192, September 1962.

\bibitem{kotz2000continuous}
N.L. Johnson, S.~Kotz, and N.~Balakrishnan,
\newblock {\em Continuous Univariate Distributions, vol. 1},
\newblock Wiley Series in Probability and Statistics, 1994.

\bibitem{Abramowitz}
M.~Abramowitz and I.A. Stegun,
\newblock {\em Handbook of Mathematical functions with Formulas, Graphs and
  Mathematical Tables.},
\newblock Dover Publications, 1970.

\bibitem{Kay1993}
S.M. Kay,
\newblock {\em Fundamentals of Statistical Signal Processing: Estimation
  Theory.},
\newblock Englewood Cliffs, NJ: Prentice-Hall, 1993.

\bibitem{Jonkman}
J.N. Jonkman, P.D. Gerard, and W.H. Swallow,
\newblock ``Estimating probabilities under the three-parameter gamma
  distribution using composite sampling,''
\newblock {\em Computational Statistics \& Data Analysis}, vol. 53, no. 4, pp.
  1099--1109, February 2009.

\bibitem{hurlin2013}
C.~Hurlin,
\newblock ``Maximum likelihood estimation,''
  \url{http://www.univ-orleans.fr/deg/masters/ESA/CH/Chapter2_MLE.pdf}, 2013.

\bibitem{scholz1985maximum}
F.W. Scholz,
\newblock ``Maximum likelihood estimation,''
\newblock {\em Encyclopedia of statistical sciences}, 1985.

\bibitem{cressie1986}
N.A.C. Cressie and H.J. Whitford,
\newblock ``How to use the two sample t-test,''
\newblock {\em Biometrical Journal}, vol. 28, no. 2, pp. 131--148, 1986.

\bibitem{wendorf2004}
C.~Wendorf,
\newblock ``Manuals for univariate and multivariate statistics,''
\newblock {\em Stevens Point, WI: University of Wisconsin}, 2004.

\bibitem{rakotomalala2013}
R.~Rakotomalala,
\newblock ``Comparaison de populations: Tests param{\'e}triques,''
\newblock {\em Bartlett test, Version}, vol. 1, pp. 27--29, 2013.

\bibitem{johnson2013}
V.E. Johnson,
\newblock ``Revised standards for statistical evidence,''
\newblock {\em Proceedings of the National Academy of Sciences}, vol. 110, no.
  48, pp. 19313--19317, 2013.

\bibitem{wiltgen2008automatic}
M~Wiltgen, A~Gerger, C~Wagner, J~Smolle, et~al.,
\newblock ``Automatic identification of diagnostic significant regions in
  confocal laser scanning microscopy of melanocytic skin tumors,''
\newblock {\em Methods of Information in Medicine}, vol. 47, no. 1, pp. 14--25,
  2008.

\end{thebibliography}
\end{document}